# Effect of wall roughness on heat transfer in supercritical water flow

Piyush Mani Tripathi, Saptarshi Basu

Department of Mechanical Engineering, Indian Institute of Science, Bangalore 560012 India

## ABSTRACT

This paper discusses the numerical investigation of the wall roughness effect on the supercritical water flow susceptible to heat transfer deterioration (HTD). The simulation was carried in the vertical circular pipe using SST k-ω turbulence model for different sets of heat flux (220kW/$m^2$ & 1810kW/$m^2$) and mass flow rate (0.0106kg/s &0.022kg/s) at a maximum pressure of 25.3MPa. The presence of roughness was incorporated as a uniform sand-grain roughness on the heated wall. As a result, HTD recuperated gradually as the roughness height ($K_s$) was increased. The mitigation of HTD is a direct consequence of the increase in turbulent kinetic energy (TKE). In contrast, the delay in the onset of HTD is due to the decrease in the dominant forces (buoyancy or acceleration effects) responsible for HTD occurrence. An equivalent thermal resistance model was proposed to elucidate the same. Additionally, the heat transfer coefficient reduces ($HTD_n$) near the outlet at higher roughness values where HTD has completely recovered. This is attributed to the high specific heat value around the pseudocritical temperature. In the end, the efficacy of the roughness was analyzed using two methods: entropy generation and thermal performance factor. Both techniques suggest that the use of rough pipe is beneficial. However, the first method recommends using a $K_s$ value greater than the critical roughness height ($K_{sc}$): as the total entropy generation was found to peak at $K_{sc}$ because of the thermal conductivity variation with temperature. The study revealed high sensitivity of maximum wall temperature and HTD onset to roughness presence. In addition, contrary to usual HTD, a new type of heat transfer impairment ($HTD_n$) can occur without any loss of TKE.

## 1. INTRODUCTION

The critical point is a thermodynamic state on the saturation curve, which marks the end of the liquid-vapor phase transition. In other words, there is no apparent phase boundary beyond a critical point; therefore, liquid and vapor cannot be differentiated further. Fluids at a pressure higher than the critical pressure are known as supercritical fluids (SF), and the same are extensively used in power and refrigeration cycles for better efficiency.

## NOMENCLATURE

| | | | |
|---|---|---|---|
| x | Axial distance from the inlet (m) | Subscript | |
| r | Radial distance from the axis (mm) | in | Inlet |
| dx | Axial grid size (mm) | out | Outlet |
| dr | Radial grid size(mm) | w | Wall |
| $y^+$ | Non-dimensional wall distance | pc | Pseudocritical |
| R, D | Radius, Diameter (mm) | b | Bulk |
| Re | Reynold number | t | turbulent |
| Nu | Nusselt number | r | reference |
| s | Entropy (J/kgK) | e | effective |
| ρ | Density (kg / $m^3$) | nm | Near wall |
| U | Average velocity (m/s) | | |
| T | Temperature (K) | | |
| q | Heat flux (kW/$m^2$) | | |
| G | Mass flux (kg/s) | | |
| λ | Thermal conductivity (W/mK) | | |
| $C_p$ | Specific heat capacity (J/kgK) | | |
| h | Heat transfer coefficient (W/$m^2$K) | | |
| TKE (k) | Turbulent Kinetic energy ($m^2/s^2$) | | |
| ε | Turbulent dissipation rate ($m^2/s^2$) | | |
| η | Thermal performance factor | | |
| $f$ | Friction factor | | |
| α | Thermal diffusivity ($m^2$/s) | | |
| $K_s$ | Roughness height (μm) | | |
| $K_s^*$ | Dimensionless roughness height ($K_s$/R) | | |

The pressure-temperature diagram beyond the critical point manifests a curve such that it marks two regions of SF with significant property variation. This curve is called a pseudo-critical line or widom line, and it demarcates the SF into the liquid-like and vapor-like region shown in Figure 1 as the shaded region[1]. However, the exact location of pseudo-critical line can be governed by different definitions such as line connecting inflection point of density at a given pressure or the locus of states with maximum specific heats at a given pressure.

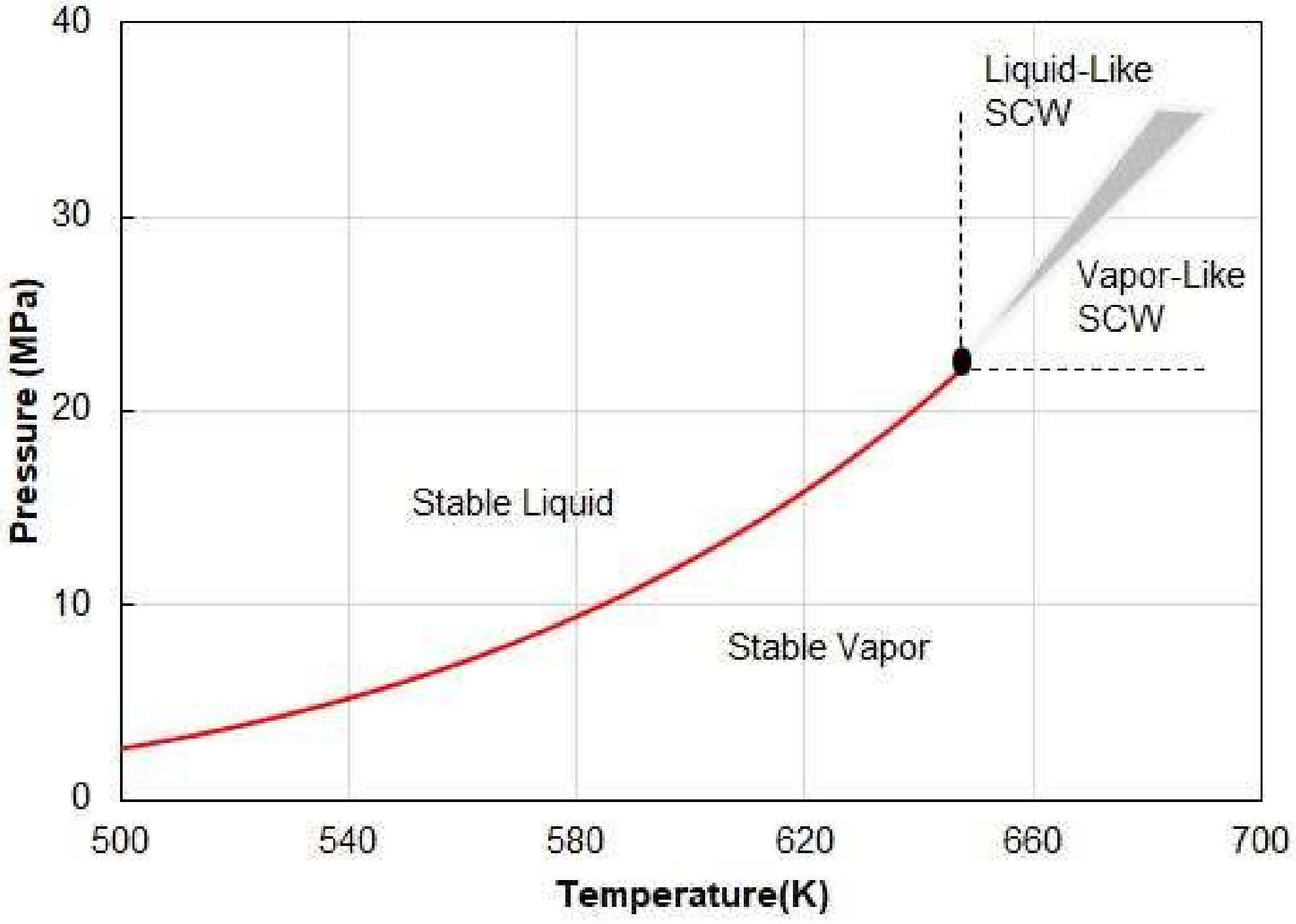

**Figure 1. Pressure Vs. Temperature (P-T) for water in the supercritical region specifying the two types of SF observed.**

SF, like supercritical water and carbon dioxide, caught the interest of researchers in the second half of the 20$^{th}$ century due to growth in the powerplant sector. Meanwhile, various experimental studies [2–5] had been carried out to investigate heat transfer characteristic of supercritical fluids, mainly in the smooth circular pipe. The results showed a significant deviation from what one would expect for a subcritical fluid. An important feature common in all the experimental observations was either heat transfer deterioration (HTD) or heat transfer enhancement (HTE). These phenomena were ascribed to thermo-physical properties variation of the supercritical fluid at the pseudo-critical region. Different properties of water as a function of temperature at pressure 25.3MPa are displayed in Figure 2.

Shitsman's[2] experimental results were first to highlight the HTD occurrence, that are identified by the peak in the wall temperature that gradually flattens as heat transfer recovers. Surprisingly for specific flow conditions, wall temperature can have two successive peaks. Initially, several studies[2,4] conjectured the occurrence of HTD to a phenomenon similar to boiling in the subcritical phase change process. However, flow alteration due to buoyancy forces was coined later to explain these observations. Shitsman[6] proposed that the transverse turbulent velocity fluctuation might reduce due to forces generated by radial density variation and the same result in HTD. Later, Hall and Jackson's studies [7–10] concluded that buoyancy forces influence the velocity profile in such a way that it reduces the turbulence level near the wall that eventually results in heat transfer impairment. Jackson[10] has summarized this idea comprehensively in a semi-empirical modal, which is supported by various experiments and simulations. Nevertheless, a different kind of HTD independent of buoyancy was reported [11,12], and it was imputed to acceleration effect emanating out of density variation.

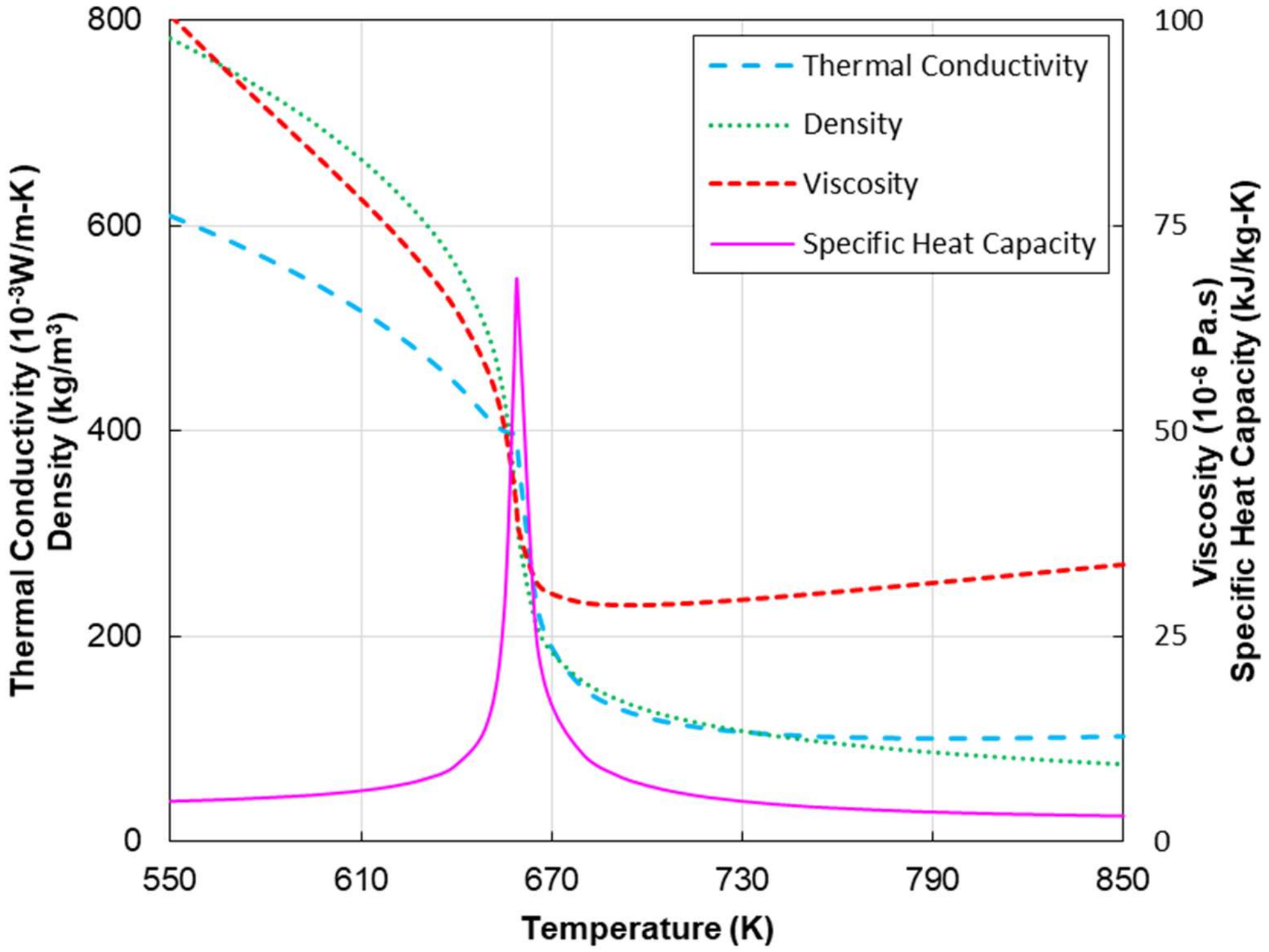

**Figure 2. Properties of water at 25.3MPa (NIST miniREFPROP) as a function of temperature near the pseudocritical point.**

Further, numerical studies [12–18] had investigated different flow conditions corresponding to previous experimental setups and had shown consistency with the result by predicting HTD and HTE. Although these works have demonstrated the capability to predict the wall temperature maxima, an exact replication of the experimental trend is yet to be established. All these investigations have found either buoyancy forces or acceleration effects to be the main reason for HTD. Moreover, these studies have shown alteration of the flow field in terms of velocity distribution, which leads to loss of turbulence kinetic energy that eventually impairs heat transfer.

However, the effect of the wall roughness on SF flows remains largely unexplored as all the studies, either experimental or numerical, have been mainly for smooth circular pipes. Roughness has an inherent presence on all the pipe surfaces, as nothing is perfectly smooth; this, by default, renders surface roughness as one of the natural flow boundary conditions. As a result, there is a high possibility that a given experimental data can not be reproduced if the surface texture of the original test section is not specified. The same discrepancy can be passed on to the HTD onset criteria and heat transfer coefficient expressions. Furthermore, for a high Reynolds number flow, even a slight roughness can be significant. For example, recent studies[19–21] on the roughness effect in the turbulent flow showed that it alters the fluid flow characteristic near the wall, affecting the heat transfer behavior. Recently there have been attempts[22,23] to unearth surface roughness effect on SF (methane) flow, but either they are very brief or based on artificial roughness (that was induced due to rib height and its pitch). Hence, understanding the detailed impact of natural roughness on SF flow is much desirable, and the same has been addressed in the present study for supercritical water.

Moreover, both types of HTD (due to buoyancy and acceleration) have been investigated in the proposed research to make it more inclusive. It is worth pointing out that the authors have investigated similar flow conditions in their previous work[24], but that was regarding the two-phase aspect of the SF flows. In other words, the present work is entirely independent of the previous study as both the research are disparate. This paper explores the roughness effect on supercritical water flow, whereas the previous study attempts to establish an analogy of a subcritical flow boiling to SF heat transfer.

In the end, the effectiveness of roughness is judged by conventional and unconventional approaches: thermal performance factor and the second law of thermodynamics, respectively. The former methods[25] do that by accounting for heat transfer gain and increased pressure loss due to roughness presence. Whereas the latter approach is inspired by A. Bejan's [26,27] work on convective heat transfer, which has been further extended for turbulent flows[28,29] and circular pipe flows[30]. Moreover, this analysis employs entropy generation calculation to analyze the roughness influence from the second law perspective.

# 2. METHODOLOGY

## 2.1 Governing equations

The primary governing equation of mass, momentum, and energy for a steady-state case is numerically solved using the commercial software (ANSYS FLUENT). The circular pipe flow is considered axis-symmetric flow, and governing equations are written for the radial and axial direction in cylindrical coordinates as follows:

Continuity equation:

$$\frac{1}{r}\left\{\frac{\partial}{\partial x}(r\rho u)+\frac{\partial}{\partial r}(r\rho v)\right\}=0 \tag{1}$$

Axial Momentum equation:

$$\begin{aligned}\frac{1}{r}\left\{\frac{\partial}{\partial x}(r\rho u^2)+\frac{\partial}{\partial r}(r\rho uv)\right\} &= -\frac{\partial P}{\partial x}-\rho g+\frac{1}{r}\frac{\partial}{\partial x}\left\{2r\mu_e\left(\frac{\partial u}{\partial x}-\frac{1}{3}(\nabla.\vec{u})\right)\right\} \\ &\quad +\frac{1}{r}\frac{\partial}{\partial r}\left\{r\mu_e\left(\frac{\partial u}{\partial r}+\frac{\partial v}{\partial x}\right)\right\}\end{aligned} \tag{2}$$

Radial Momentum direction:

$$\begin{aligned}\frac{1}{r}\left\{\frac{\partial}{\partial x}(r\rho vu)+\frac{\partial}{\partial r}(r\rho v^2)\right\} &= -\frac{\partial P}{\partial r}+\frac{1}{r}\frac{\partial}{\partial r}\left\{2r\mu_e\left(\frac{\partial v}{\partial r}-\frac{1}{3}(\nabla.\vec{u})\right)\right\} \\ &\quad +\frac{1}{r}\frac{\partial}{\partial x}\left\{r\mu_e\left(\frac{\partial u}{\partial r}+\frac{\partial v}{\partial x}\right)\right\}-2\mu\frac{v}{r^2}+\frac{2}{3}\frac{\mu}{r}(\nabla.\vec{u})\end{aligned} \tag{3}$$

$$\nabla.\vec{u}=\frac{\partial u}{\partial x}+\frac{\partial v}{\partial r}+\frac{v}{r} \tag{4}$$

Here $u, v$ are axial and radial velocity, respectively, $\rho$ is the density and $\mu_e$ is the effective viscosity defined by $\mu_e = \mu + \mu_t$ . Here $\mu_t$ is the turbulent viscosity determined by the turbulence model.

Energy equation:

$$\frac{1}{r}\left\{\frac{\partial}{\partial x}(r\rho u H)+\frac{\partial}{\partial r}(r\rho v H)\right\} = \frac{1}{r}\left\{\frac{\partial}{\partial x}\left[rC_p\left(\frac{\mu}{Pr}+\frac{\mu_t}{Pr_t}\right)\frac{\partial T}{\partial x}\right]+\frac{\partial}{\partial r}\left[rC_p\left(\frac{\mu}{Pr}+\frac{\mu_t}{Pr_t}\right)\frac{\partial T}{\partial r}\right]\right\} + D_{visc} \quad (5)$$

Here H is the specific enthalpy, Pr is the molecular Prandtl and $D_{visc}$ is the viscous dissipation term. $Pr_t$ is the turbulent Prandtl number which is equal to $0.85$ for this calculation. Further, the SST k-ω turbulence model is used for this study, which has the advantage that it subsumes the qualities of the k-ω model in the near-wall region and the k-ε model in the free stream flow. It does so by multiplying both the models with blending functions and adding them together. The blending function is equal to unity in the region near the wall and zeroes away from the surface. So, it activates k-ω near the wall and k-ε in the far-field. It is recommended to use SST k-ω because it predicts the result accurately at the wall as well as in the free stream compared to other models.

Furthermore, Roughness effects have been incorporated in the flow as closely packed uniform sand-grain roughness on the wall, and the same has been included in the numerical model by modifying the law-of-the-wall for mean velocity. For turbulent flow in rough pipes, the velocity profile slope is the same as a smooth pipe but has a different intercept depending on the roughness. The value of the intercept depends on two parameters: Roughness height ($K_s$) and the Roughness Constant ($C_s$). In the present study, the intercept formula proposed by Cebeci and Bradshaw[31] based on the Nikuradse experiment (he generated resistance data for rough pipe with tightly-packed uniform sand-grain roughness) has been used. For uniform sand-grain roughness, $K_s$ is simply the height of the grain, whereas $C_s$ depends on the roughness characteristics. When the k-ε model is used to reproduce the experimental result of Nikuradse, $C_s$ value comes out to be $0.5$, and the same has been used in the computational model.

In the end, the effectiveness has been investigated based on two different approaches:

The thermal performance factor (η) is proposed to weigh the benefit of increased heat transfer against the additional pressure loss induced due to roughness presence[25]. It has been defined in the following way:

$$\eta = \frac{Nu/Nu_r}{(f/f_r)^{1/3}} \quad (6)$$

Here, subscript 'r' in equation (6) corresponds to reference, which is usually the smooth pipe. Moreover, the numerator accounts for gain in convective heat transfer, whereas the denominator gauges the extra loss in the pressure due to roughness inclusion.

For the second law analysis, entropy generation is computed. The entropy per unit mass (s) has a balanced equation for a steady-state given as following[29].

$$\rho(\vec{u}.\nabla s) = \nabla.\left(\frac{q}{T}\right) + \frac{\emptyset}{T} + \frac{\emptyset_\theta}{T^2} \quad (7)$$

Where $\frac{\emptyset}{T}$ and $\frac{\emptyset_\theta}{T^2}$ are the per unit volume entropy generation ($\dot{S}$) terms. The first term ( $\frac{\emptyset}{T}$ ) describes the entropy production due to dissipation of kinetic energy and for turbulent flow has two parts: $\dot{S}_{U1}$ is due to viscous dissipation which is defined in cylindrical coordinate as[30]:

$$\dot{S}_{U1} = \frac{\mu}{T}\left[2\left\{\left(\frac{\partial u}{\partial x}\right)^2 + \left(\frac{\partial v}{\partial r}\right)^2 + \left(\frac{v}{r}\right)^2\right\} + \left(\frac{\partial u}{\partial r} + \frac{\partial v}{\partial x}\right)^2\right] \quad (8)$$

$\dot{S}_{U2}$ is caused by turbulent dissipation ( dissipation due to turbulent fluctuations of velocity), which is expressed as following[28,29]

$$\dot{S}_{U2} = \frac{\rho\varepsilon}{T} \quad (9)$$

The second term ( $\frac{\emptyset_\theta}{T^2}$ ) describes the entropy production term due to heat transfer and for turbulent flow has two parts: $\dot{S}_{T1}$ is entropy production term due to mean temperature gradients and described in cylindrical coordinate as[30]:

$$\dot{S}_{T1} = \frac{\lambda}{T^2}\left[\left(\frac{\partial T}{\partial x}\right)^2 + \left(\frac{\partial T}{\partial r}\right)^2\right] \quad (10)$$

$\dot{S}_{T2}$ is the entropy generation term due to gradients of temperature fluctuations and defined as following [28,29].

$$\dot{S}_{T2} = \frac{\alpha_t}{\alpha}\dot{S}_{T1} \quad (11)$$

## 2.2 Physical and Numerical model.

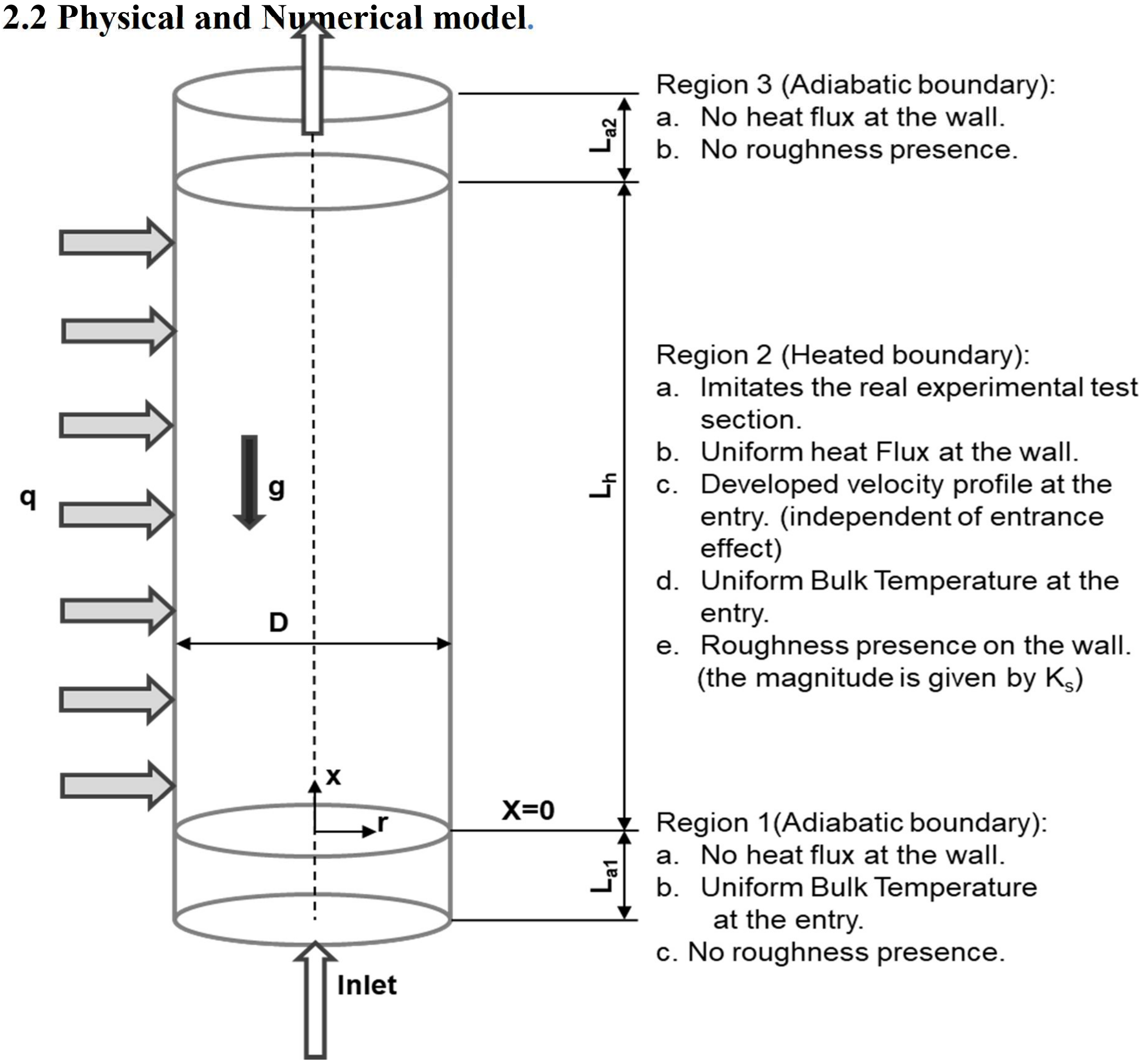

**Figure 3. Schematic diagram of flow geometry used for computation purpose and the geometric details of the same is given in table 1.**

In the present study, the circular flow domain, as displayed in figure 3, has been simulated, and the geometric details vary for different cases based on the specification given in table 1. The flow geometry used is a vertical circular pipe, and the same has three distinct regions along the axis for computational purposes. The first region consists of an adiabatic wall from the inlet in the flow direction to ensure that a velocity profile is developed so that the entrance effect is negligible at the next region's entry. Then follows the second region, which is heated with uniform heat flux at the wall, and its length is similar to the experimental section. Besides, the roughness is incorporated at the boundary in this region, as mentioned in figure 3. Next, there is a third region with the adiabatic wall up to the outlet. The last part ensures that there is no backward flow, resulting in a stable computational system.

Furthermore, like usual flow, a no-slip boundary condition exists at the wall; thus, a high radial gradient of flow variables is expected near the wall. So, to correctly resolve this region and accurately predict the results, it is recommended to have a value of less than 1 for $y^+$ of the first element near the wall. The same has been achieved by introducing non-uniform mesh in the radial direction such that the highest mesh density is maintained in the wall vicinity. The size of the first grid near the wall is $0.001$ mm, which has ensured that the value of $y^+$ is less than $0.2$, which is way below the recommended limit. In the end, the National Institute of Standards and Technology (NIST) software miniREFPROP has been referred for the thermo-physical properties of water at various operating conditions. For all the simulations, a piecewise linear profile is used to incorporate the properties as a function of temperature for a given pressure. Incorporating properties independent of pressure is a valid assumption because of the significantly less pressure drop in the flow compared to the absolute pressure value.

## 2.3 Boundary conditions

The boundary conditions are given in Table 1 that span over all the cases and are inspired from Shitsman's[2] and Ornatskij's experiment (the flow condition has been taken from [12]). In Shitsman's setup, inlet bulk enthalpy is mentioned instead of temperature. So the NIST miniREFROP has been used to find out the inlet bulk temperature ($T_{in}$) for given enthalpy and pressure. At the inlet boundary, turbulent intensity and turbulent viscosity ratio are set as 5% and 10, respectively. Besides, mass flow inlet and pressure outlet are chosen as boundary conditions considering the fact that fluid is compressible in the flow.

Uniform heat fluxes and roughness are provided on the heated wall (region 2) only, as shown in figure 3. $K_s$ is the absolute roughness height, and the roughness constant ($C_s$) is kept unchanged at 0.5 for all the simulations. The value of $K_s$ is varied from 0 (smooth wall) to 40μm and 15μm, for the case2 (as well as case-1) and case-3, respectively, as given in table 1. In the present study, $K_s$ has also been expressed in the dimensionless form as $K_s^*$ (=$K_s/R$). The gravitational acceleration value is maintained at $9.8 m/s^2$. The pseudocritical temperature ($T_{pc}$) of water at 25.3MPa and 23.3 MPa is 386 °C (659 °K) and 378.75°C (651.75 °K), respectively.

**Table 1. Boundary conditions and geometric specification of the flow domain for different cases.**

| Parameter | Case-1 | Case-2 | Case-3 |
|---|---|---|---|
| Reference | Shitsman's experimental setup with no HTD[2] | Shitsman's experimental setup with HTD having two peaks in wall temperature[2] | Inspired from Ornatskij experimental setup with HTD[12] |
| G | 0.022 kg/s | 0.022 kg/s | 0.0106 kg/s |
| $T_{in}$ | 600 K | 578 K | 500 K |
| q | 220.8 $kW/m^2$ | 384.8 $kW/m^2$ | 1810 $kW/m^2$ |
| $P_{out}$ | 23.3 MPa | 25.3 MPa | 25.3 MPa |
| $Re_{in}$ | 43129 | 38885 | 36958 |
| D | 8 mm | 8 mm | 3 mm |
| $K_s$ | (1,5,10,15,20 & 40)µm | (1,5,10,15,20 & 40)µm | (1,2,4,6,8 & 15)µm |
| $L_{a1}$, $L_h$ & $L_{a2}$ | (100, 1500 &100) mm | (100, 1500 &100) mm | (100, 800 &100) mm |

# 3. RESULTS AND DISCUSSION

## 3.1 Model validation with experimental results (smooth pipe) and grid independence

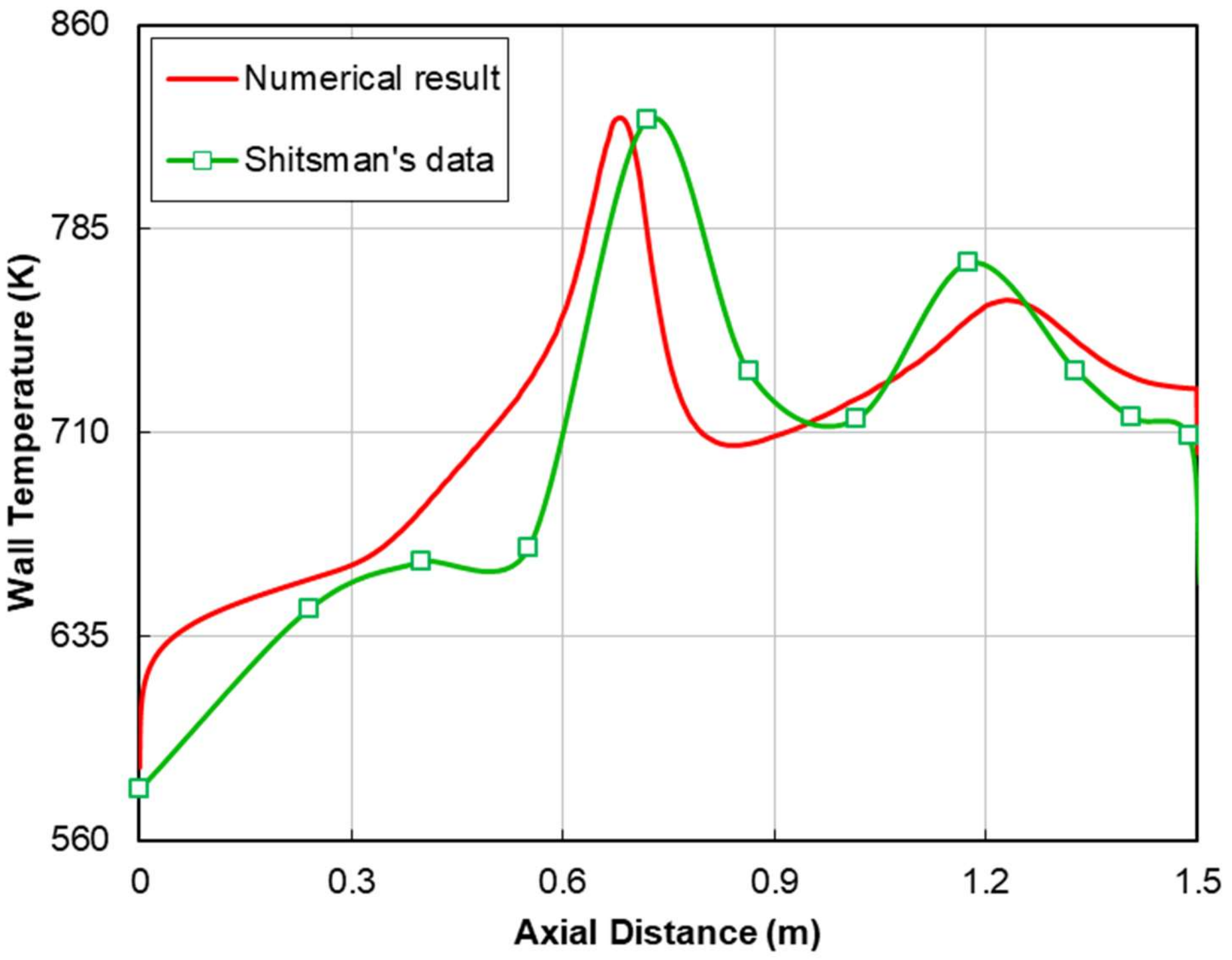

**Figure 4. Wall temperature comparison of numerical result and experimental data for case-2 to validate the computational model.**

Figure 4 shows a juxtaposition of the numerical model result and Shitsman's experimental data[2] for a smooth circular pipe. The simulation result is in good agreement with the experiment and reasonably predicts the two peaks in wall temperature that signify HTD. Although there is a minor discrepancy in the magnitude and location of the second peak, the computation model is competent enough to predict the localized heat transfer impairment, and the same has been used throughout the analysis for all the cases.

In addition, the results discussed in the paper are grid independent for all the cases, and in the process, the mesh was refined in the radial and the axial direction both. In this section, the grid-independent study of case-1 and case-3 have been presented; however, a similar process was adopted for case-3. For both the cases (1&3), initially, there were 200000 nodes in the coarse grid, and then it was refined subsequently to finer mesh. Although mesh is improved radially, the first grid point near the wall remains unchanged in all the simulations to achieve the desired $y^+$ value.

The wall temperature variation plot in figure 5(a) for different mesh specifications shows that there is always an improvement at the peaks, and the rest temperature profile remains almost the same. However, improvement in the peak temperature value decreases with mesh refinement, as manifested in Figure 5(a). It conveys that the grid independence at the peaks is relatively slow with respect to mesh refinement. On the contrary, case-1 approaches grid independence as displayed in figure 5(b) for the same refinement process. Furthermore, one can anticipate this because of the high thermal gradient at the wall for case-2 compared to case-1, as the former suffers from HTD. For a better perspective, the plot of thermal gradient variation for both cases has been shown in Figure 6(a). It is visible that the thermal gradient near the wall in case-2 is nearly four times of case-1, at the location of the first HTD occurrence.

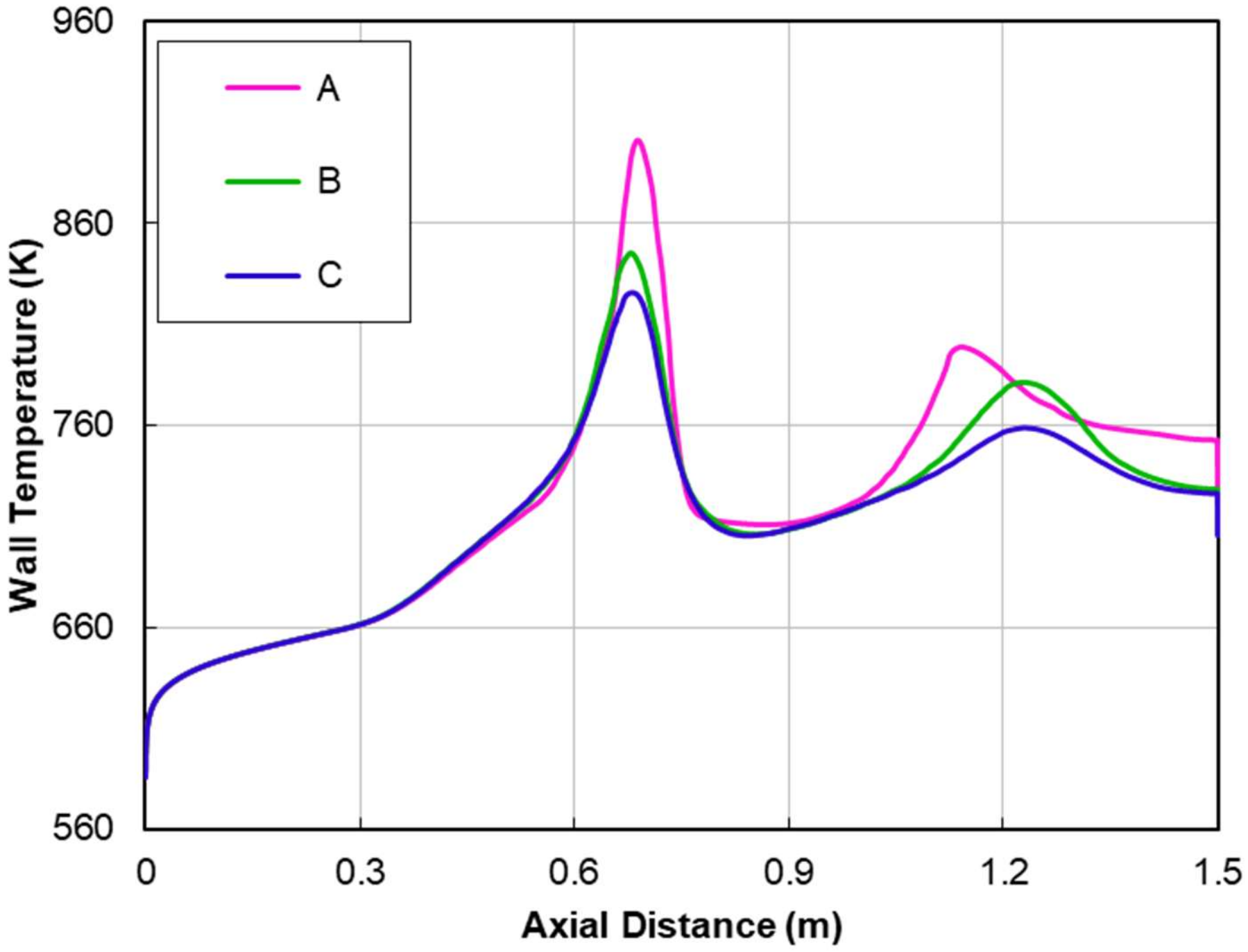

**(a) A= 200000 nodes, B=2A and C=2B**

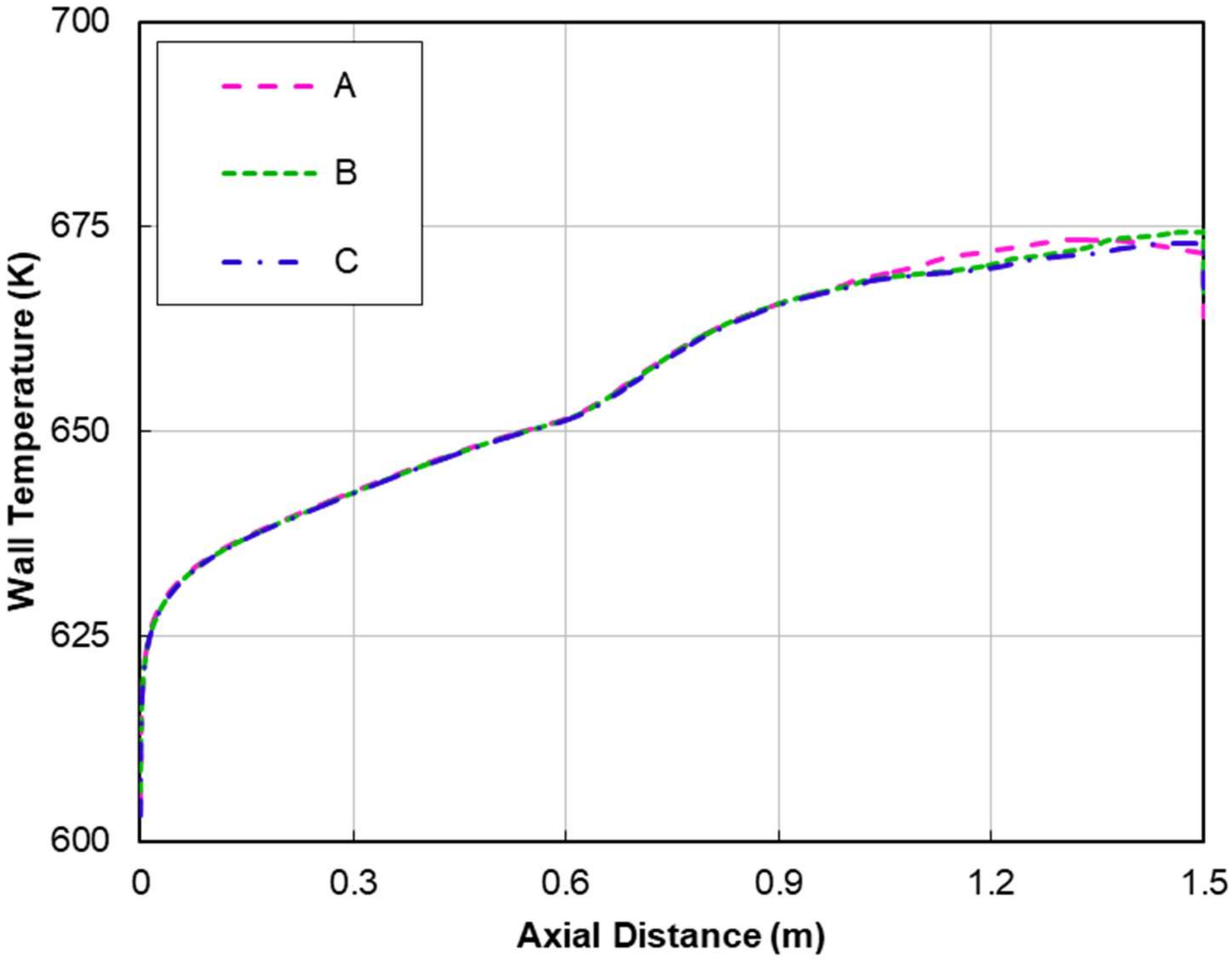

**(b) A= 200000 nodes, B=2A and C=2B**

**Figure 5. Axial variation of wall temperature for the different number of nodes in the process of grid-independent study. (a) Case-2 (b) Case-1**

Further, it has been observed that the wall temperature is insensitive to axial mesh refinement, and even this has been mentioned as a grid independence study in some of the previous studies. In other words, results are unchanged when the mesh is refined axially, keeping the radial grid structure the same at each simulation. The same has been demonstrated for case-2 with an initial axial mesh element (dx) size of 0.6 mm corresponding to 200000 nodes and then decreased dx by a factor of two keeping everything the same. Figure 6(b) demonstrates the insensitivity of wall temperature to axial refinement for case-2. Based on the above results and computational cost, mesh having 440000 and 120000 grid points for case-2 (as well as case-1) and case-3, respectively, have been used in the present analysis.

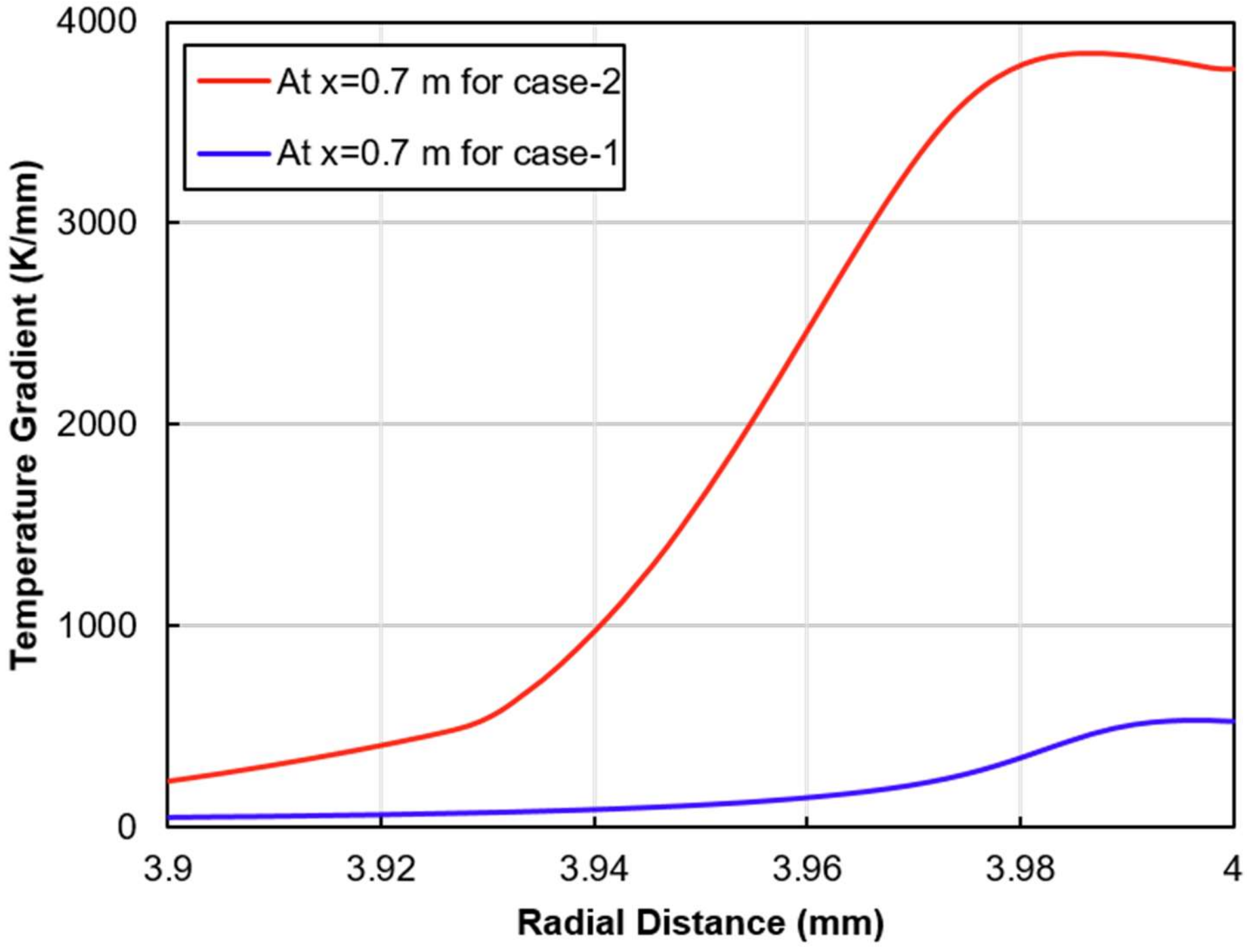

**(a)**

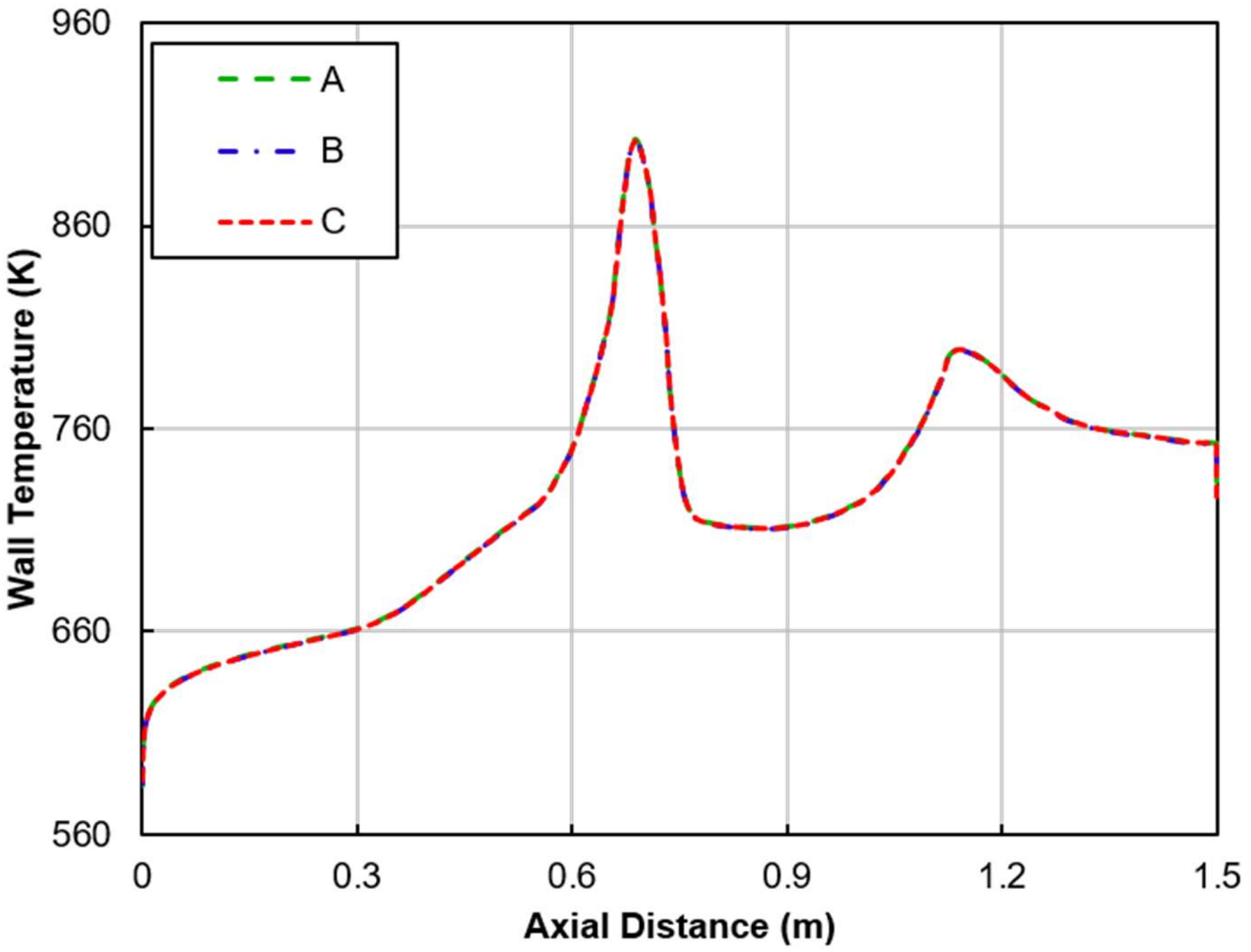

**(b) A= 200000 nodes, B=2A and C=2B**

**Figure 6. (a) Comparison of radial temperature gradient near the wall for case-1 & 2 corresponding to the axial location (x=0.7m) of first HTD occurrence for case-2. (b) Axial variation of wall temperature as a result of only axial refinement of mesh for case-2.**

### 3.2 Simulation with roughness

All the simulations were performed on the same numerical setup used for validation in section 3.1, and the only new addition is the incorporation of roughness on the wall of the heated section (region 2), as shown in figure 3. For better clarity, this section is subdivided into two subparts, where each discusses the results for case-2 and case-3. The heat transfer coefficient (h) was enumerated in the following section using the definition given in equations (12) & (13).

$$T_b = \frac{\int_0^R 2\pi T C_p \, \rho u r \, dr}{\int_0^R 2\pi C_p \, \rho u r \, dr} \tag{12}$$

$$h = \frac{q}{T_w - T_b} \tag{13}$$

Furthermore, the cases discussed in this section experience HTD, which is a direct consequence of the decline in turbulence. Since the convective heat transfer involves bulk motion of fluid and any interference to fluid movement will affect the heat transfer characteristic. The TKE judges the turbulence level in the flow, and any loss in the same will be predicted by a plunge in TKE value marked by flattening of the velocity profile. So, velocity and TKE profile have been analyzed to assess the roughness effect.

### 3.2.1 Case-2 (HTD due to buoyancy)

The HTD, which was apparent in a smooth pipe ($K_s$ or $K_s^* = 0$), vanished gradually as the roughness height increases, eventually resulting in flattening of the wall temperature peaks for $K_s \geq 15\mu m$ (or $K_s^* = 37.5 \times 10^{-4}$) as shown in figure 7(a). It is because of the rise in the flow turbulence level induced by the roughness presence. The same will weaken the HTD and finally restore normal heat transfer behavior. Besides, the onset of HTD is delayed with increases in the $K_s$ magnitude, where the axial location of the temperature maximum propagates further downstream, as displayed in figure 7(a). This aspect of roughness presence does not seem that obvious and requires further investigation (discussed in section 3.3).

Figure 7(b) displays the axial variation of h value for different $K_s^*$ values. The initial drop in the h value near the outlet is a result of the thermal entrance effect, where the thermal boundary layer is in the developing phase. Further downstream, the flow starts experiencing the influence of HTD. Moreover, h experiences a minimum at axial positions, which are the same as the HTD locations for the given $K_s^*$. The second minimum is shallower than the first one, which corroborates the intensity of heat transfer impairment experienced at corresponding axial locations. The significant impact of the roughness presence in the bulk of the flow is only evident after a particular value of roughness height ($K_{st} \sim 10\mu m$ or $K_{st}^* \sim 25 \times 10^{-4}$) as shown in figure 7(b). Also, as roughness increases, the heat transfer coefficient improves for all axial locations, and HTD recuperates. However, the magnitude of h decreases slightly near the outlet after having a monotonous increase for $K_s^* = 50 \times 10^{-4}$ (or $K_s = 20\mu m$).

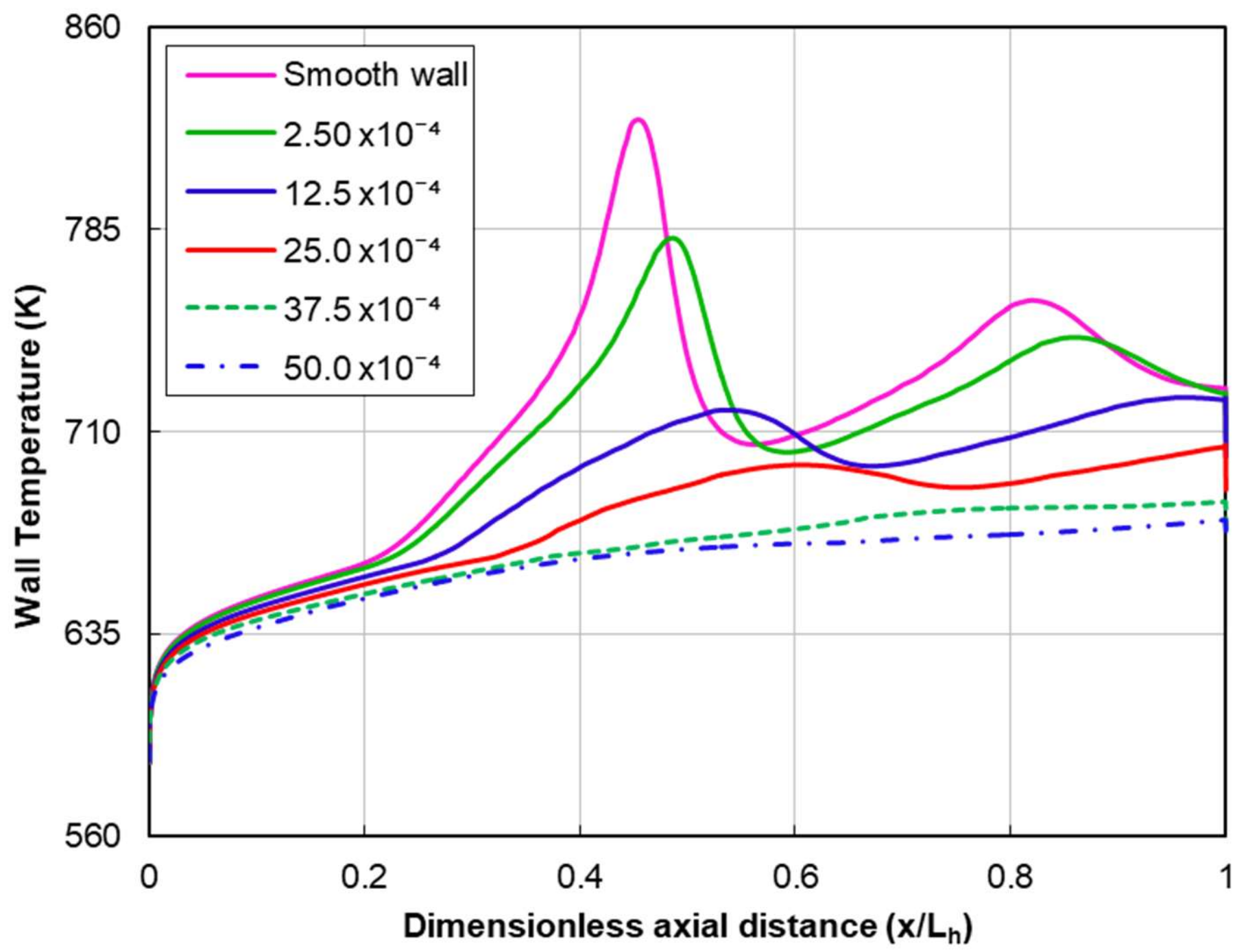

(a)

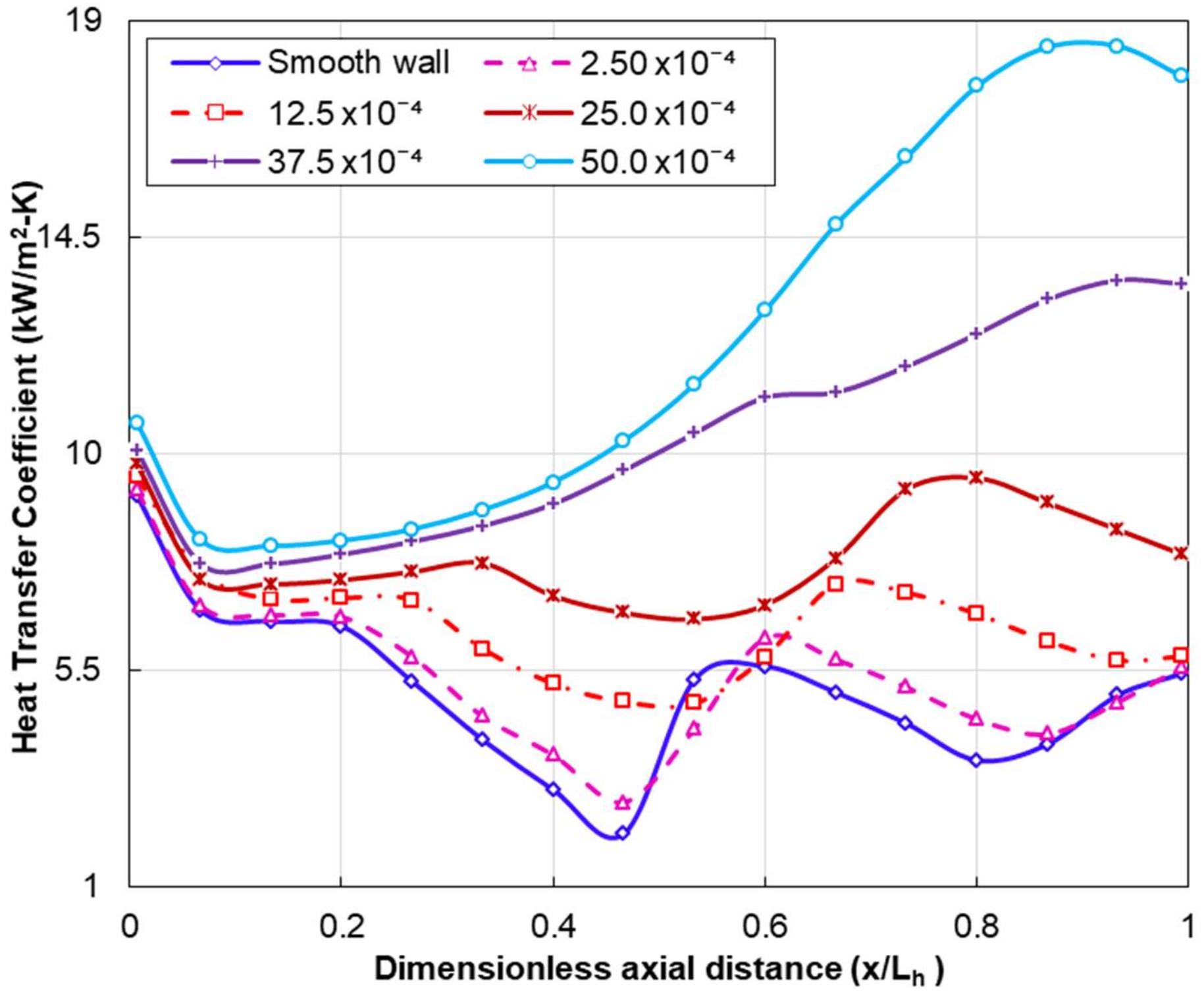

(b)

**Figure 7. Results of simulation with roughness for case-2 (a) Wall temperature profile for various $K_S^*$. (b) Axial variation of heat transfer coefficient for different $K_S^*$. For both the plot, numbers in the legend show the $K_S^*$ magnitude.**

For a better perspective, velocity and TKE profile have been displayed at the same axial locations for different $K_s^*$ values in Figures 8(a) & (b). Figure 8(a) shows the radial variation of the velocity at the x=0.7m, the axial location of the first HTD occurrence in the smooth pipe. It reveals the way the usual 'M-shaped' velocity profile observed in buoyancy-dominated HTD is manipulated in the presence of roughness. In other words, during HTD occurrence, buoyancy forces generated out of radial density variation results in higher velocity in the wall neighborhood of smooth pipe. But with an increase in $K_s$ value, turbulence levels are augmented, which eventually smoothes the density variation. Consequently, the strength of buoyancy declines that leads to a reduction in the velocity near the wall. At a higher roughness value ($K_s$ = 20μm or $K_s^*$= 50 x$10^{-4}$), when HTD does not occur, velocity follows a usual turbulent flow variation where its value decreases gradually from the center to zero at the wall, as shown in figure 8(a). It occurs due to more uniformity in the temperature as well as density across the cross-section due to higher turbulence levels which will ultimately render buoyant forces insignificant.

Moreover, figure 8(b) represents the expected trend of increasing TKE with roughness at the same axial location of x=0.6m. The rise in turbulence level in the wall vicinity is visible with the inclusion of the very slight roughness ($K_s$ ~ 1μm or $K_s^*$= 2.5 x$10^{-4}$). However, this plot further bolsters the observation that the significant effect of roughness is felt on the bulk of flow only after a certain roughness value ($K_{st}^*$).

Although the effect of roughness is experienced by the bulk of flow after $K_{st}^*$, the response of wall temperature is very prominent even for the smallest roughness value ($K_s^*$ = 2.5 x$10^{-4}$ < $K_{st}$) investigated for case-2, as shown in Figure 7(a). It can be explained by the fact that for lower roughness value, the influence is confined to the neighborhood of the wall only, and the main flow remains unaware of the roughness presence. But this influence is significant enough to improve the localized heat transfer deterioration, which is substantially governed by the flow structure near the wall.

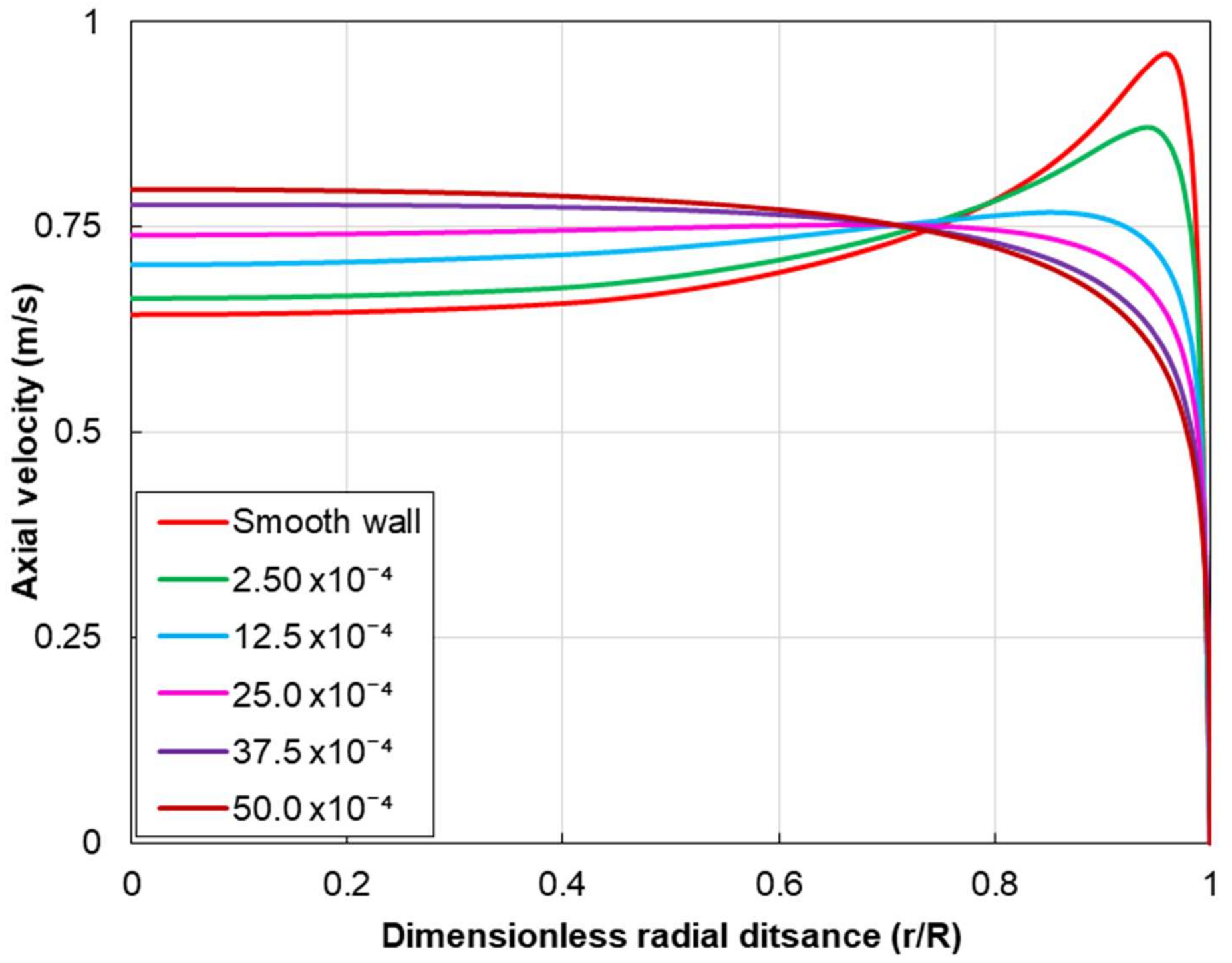

**(a)**

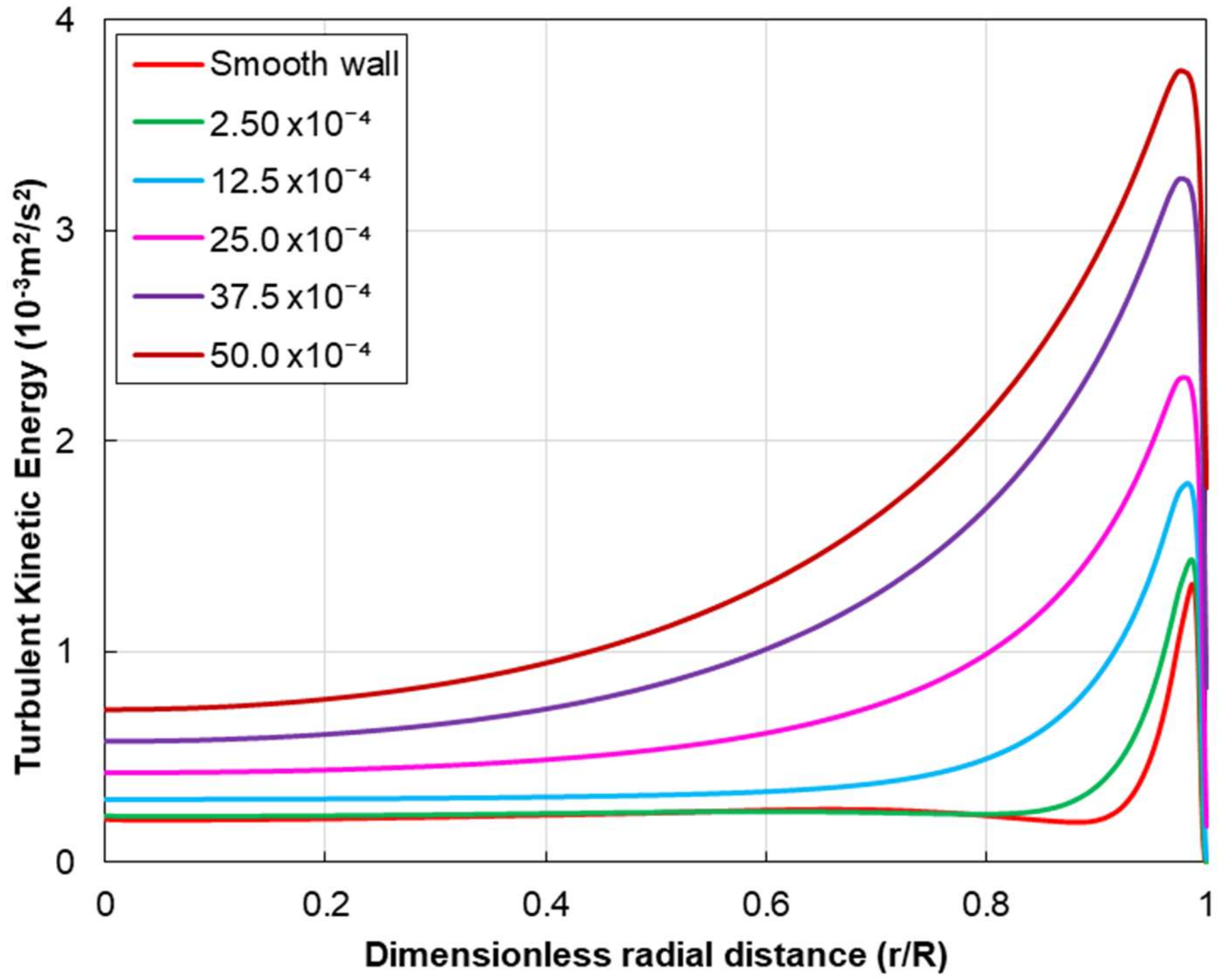

**(b)**

**Figure 8. Results of simulation with roughness for case-2 (a) Velocity profile for different $K_s^*$ value at the axial location of x=0.7m. (b) Radial variation of TKE at axial location of x=0.6m for various $K_s^*$ value. For both the plot, numbers in the legend show the $K_s^*$ magnitude.**

### 3.2.2 Case-3 (HTD due to acceleration effect)

HTD due to acceleration effects is fundamentally different from the one caused by buoyancy as the origin of the two forces responsible for heat transfer impairment are distinct. The former is generated out of axial density variation (influence on the bulk of flow), whereas the former results from the radial density disparity (concentrated near the wall). This difference is reflected in the HTD characteristic; for example, case-3 has a broad and gradual wall temperature peak, unlike case-2, which has a sharp and steep temperature maximum.

Figure 9(a) shows the decline in the intensity of HTD with increases of $K_s^*$ value, and ultimately at higher roughness ($K_s \geq 6\mu m$ or $K_s^* \geq 40 \times 10^{-4}$), it appears that heat transfer has recovered completely. The plot corroborates the finding of case-2, where the onset of HTD is delayed with increased roughness value. In other words, the axial position at which wall temperature starts rising due to the occurrence of HTD is shifted downstream with an increase in $K_s$ magnitude compared to a smooth pipe. The heat transfer coefficient (h) profile displayed in figure 9(b) also predicts the HTD and shows a gradual decline in its value. This trend is similar in shape for all the roughness values ($K_s^* < 40 \times 10^{-4}$) where HTD still exists, but the intensity of decreases in h is less severe as expected. Besides, similar to case-2, it also experiences a thermal entrance effect near the inlet, although the decline is smaller compared to case-2 because of higher velocity.

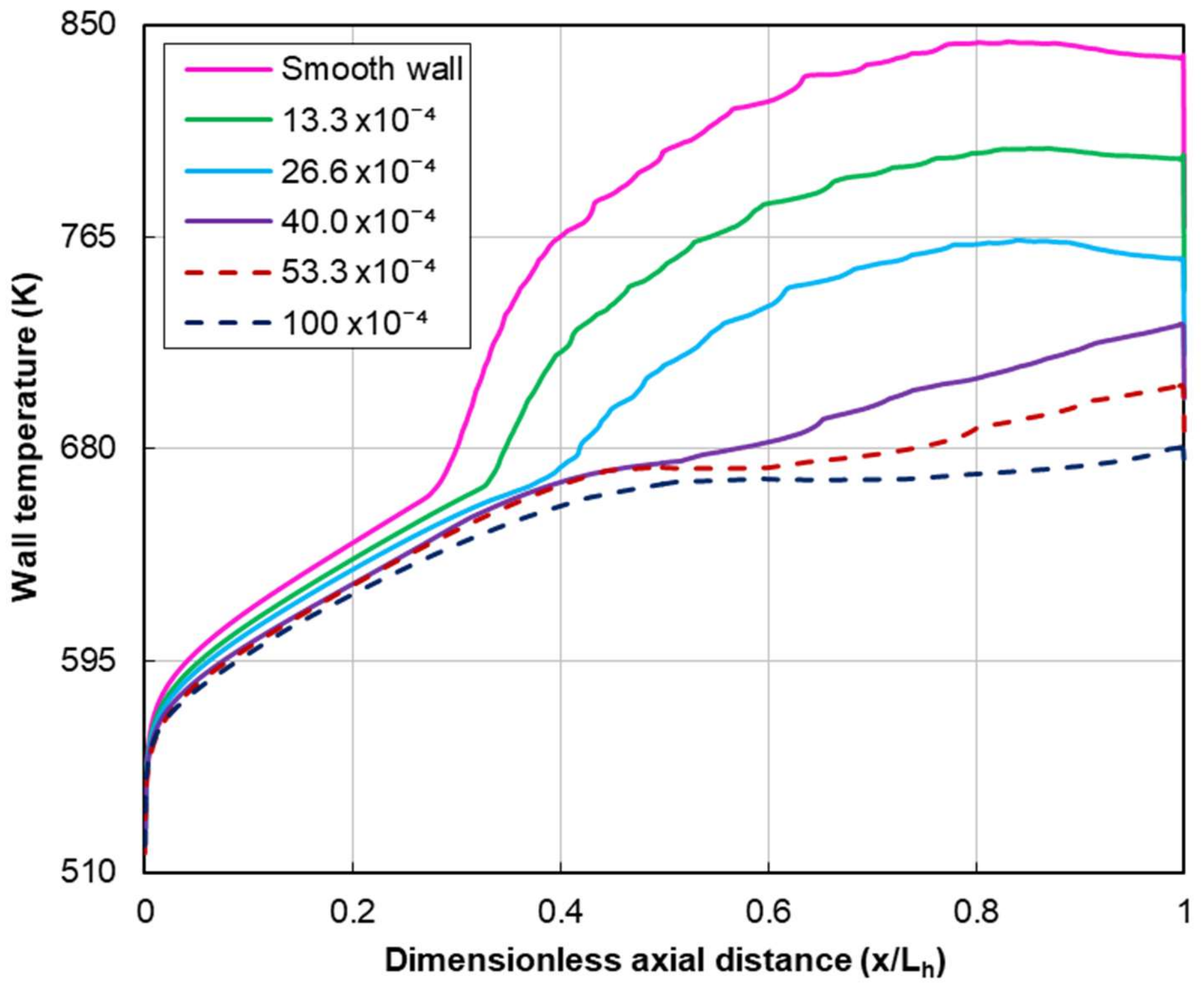

**(a)**

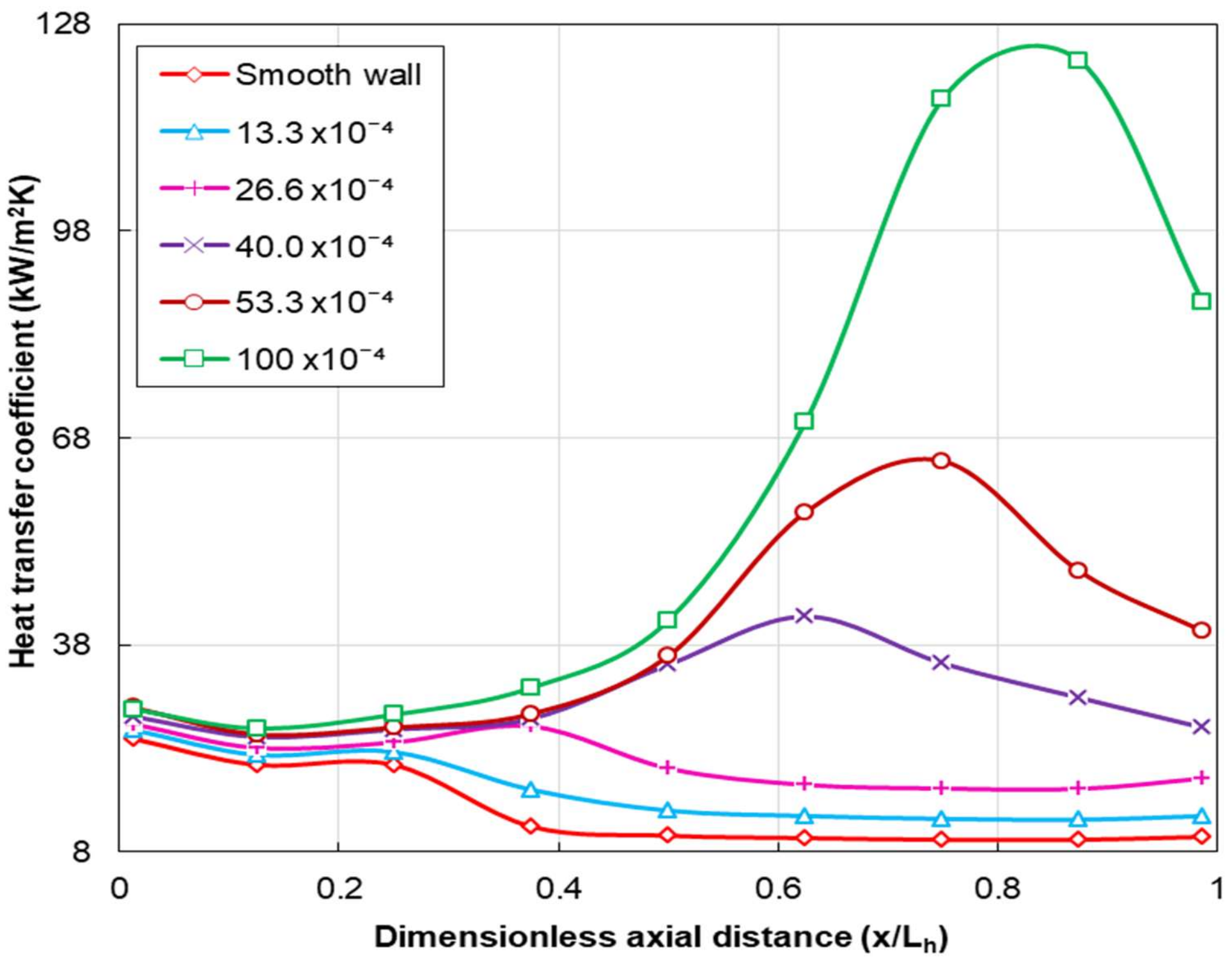

**(b)**

**Figure 9. Results of simulation with roughness for case-3 (a) Wall temperature variation for various $K_S^*$values. (b) Axial profile of heat transfer coefficient for different $K_S^*$. For both the plot, numbers in the legend show the $K_S^*$ magnitude.**

Surprisingly, even after the HTD has recovered, h experiences a drop in its value preceded by a maximum for $K_S^* \geq 40$ x$10^{-4}$. The magnitude of the peak and the subsequent decline increases with $K_S^*$. Besides, the peak shifts downstream and moves closer to the outlet with an increase in roughness value, as shown in figure 9(b). This observation points toward some form of heat transfer impairment ($HTD_n$) different from the usual HTD reported in SF flow. It is also reflected in the wall temperature profile, where it rises near the outlet for $K_S^* \geq 40$ x$10^{-4}$, as shown in figure 9(a).

In addition, similar to case-2, figure 10 (a) & (b) represent the velocity and TKE profile, respectively, for various $K_S^*$ values at the axial location of x=0.5m. As Case-3 belongs to the SF flow category where acceleration effects cause HTD; as a result, the velocity alteration is mainly concentrated in the bulk of the flow. For better understanding, figure 10(a) displays the velocity profile away from the wall for all the $K_S^*$. In this plot, the horizontal axis (r/R) is restricted to 0.9 instead of 1.

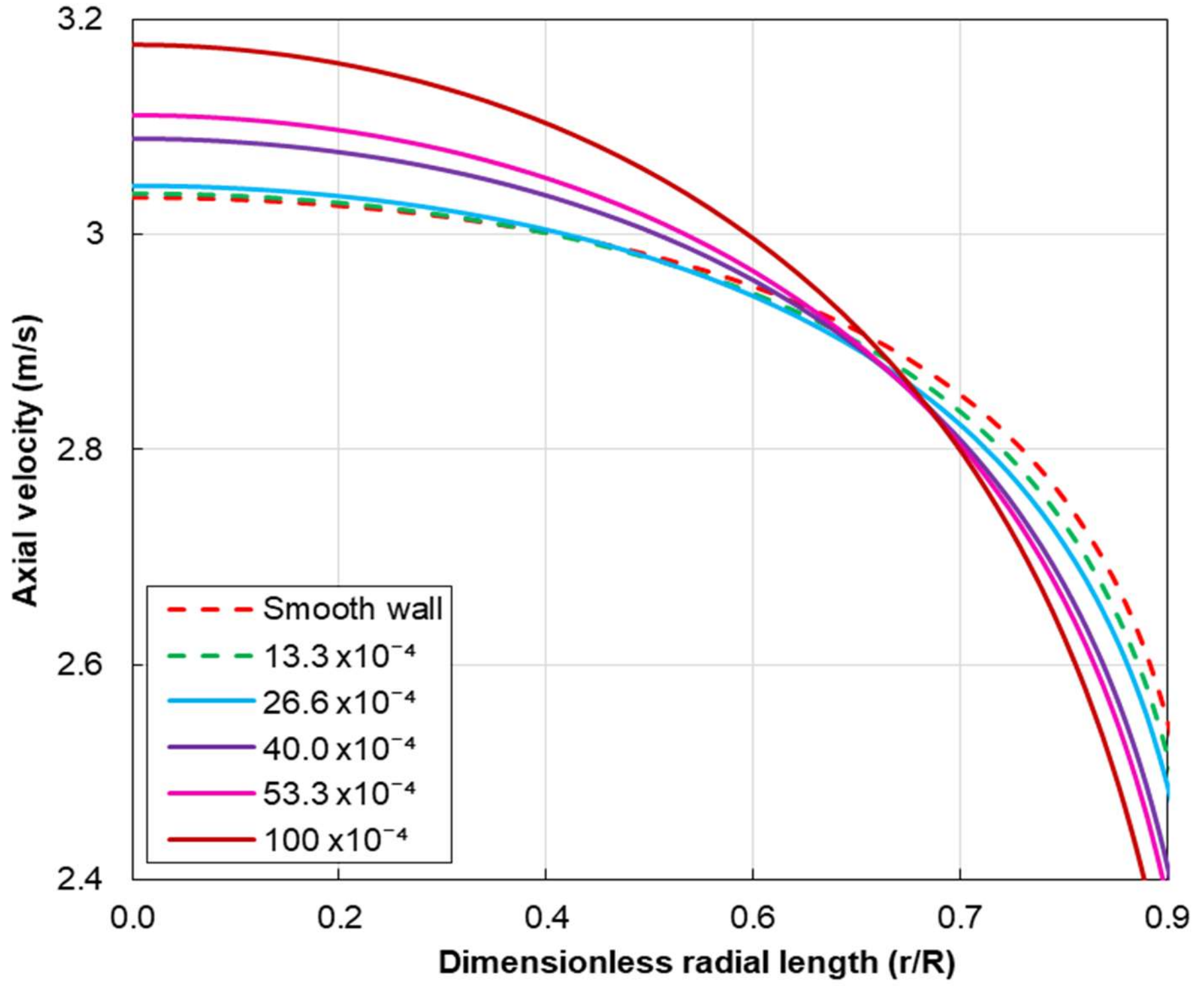

**(a)**

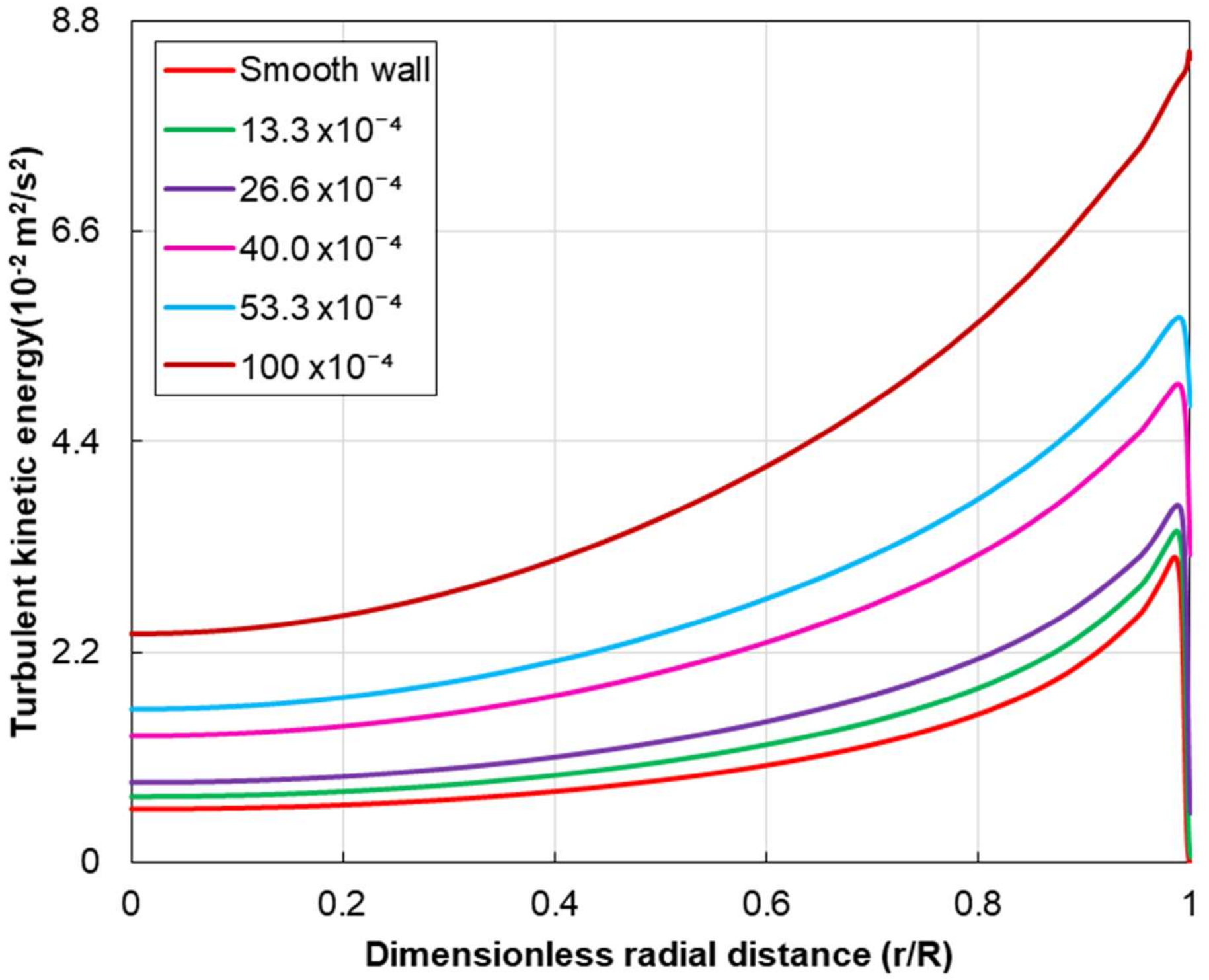

**(b)**

**Figure 10. Results of simulation with roughness for case-3 (a) Velocity profile for different $K_s^*$ values at the axial location of x=0.5m. (b) Radial variation of TKE at the axial position of x=0.5m for various $K_s^*$values. For both the plot, numbers in the legend show the $K_s^*$ magnitude.**

The plot reveals that the velocity decreases near the wall and increases in the bulk with an increase in roughness compared to the velocity profile observed in the usual HTD reported in a smooth pipe. Similar to case-2, the velocity approaches a typical turbulent flow velocity profile for a higher roughness value ($K_s$ = 15μm or $K_s^*$=100 x$10^{-4}$). This can be seen as a natural consequence of increased turbulence level that ensures more uniformity in temperature distribution in the flow. All this leads to reduced axial density variation resulting in weakening acceleration effects and ultimately rendering it insignificant. Further, figure 10(b) represents TKE variation, and similar to case-2, the significant impact of roughness is felt by the bulk of flow only beyond a certain roughness value ($K_{st}$ ~ 4μm or $K_{st}^*$ ~ 26.6 x$10^{-4}$). However, TKE rises near the wall with the introduction of very little roughness ($K_s$ = 1μm or $K_s^*$ = 6.65 x$10^{-4}$ < $K_{st}^*$) and the same is reflected in the wall temperature profile where HTD intensity decreases with the incorporation of minuscule roughness presence on the wall.

## 3.3 Roughness influence on HTD: Interpretation

### 3.3.1 Delay in the onset of HTD and its high roughness sensitivity

The intriguing and common observation is the delay in the onset of HTD compared to the smooth pipe due to roughness inclusion. This observation is not that lucid because the increase in turbulence level can not completely explain the downstream shift of the heat transfer impairment region, unlike the decline in the HTD intensity due to roughness. In other words, if there is no effect other than increased turbulence, then the wall temperature profile shape should remain the same, although its magnitude should reduce with the rise in $K_s$. For better understanding, an equivalent thermal resistance model drawing inspiration from conduction heat transfer has been proposed, and the same is shown in figure 11(a). It represents the SF flow in the form of a thermal circuit: $\lambda_{nw}$ and $\lambda_b$ represent near-wall and bulk conductivity respectively, similarly $T_w$, $T_{nw}$, and $T_c$ stand for wall temperature, near-wall temperature, and centerline temperature, respectively.

Furthermore, it has been established that a significant effect of roughness in the bulk of flow is only experienced only after the threshold roughness ($K_{st}^*$) value (~25 x$10^{-4}$for case-2 & ~ 26.6 x$10^{-4}$ for case-3). This essentially means that the bulk flow structure remains almost similar for all the roughness presence smaller than $K_{st}^*$. If we borrow this understanding into the thermal resistance model, it will ensure that the $T_c$ and $\lambda_b$ have a value similar to the smooth pipe for the roughness lying in the range 0 < $K_s^*$< $K_{st}^*$. As a result, for constant heat flux $T_{nb}$ will also remain the same for all the setup. Besides, flow in the wall vicinity is influenced even for the roughness value less than $K^*_{st}$. It means that $\lambda_{nw}$ is augmented with the inclusion of minimal roughness even if it does not cross the threshold value. If we employ the Fourier law of conduction for the constant heat flux case, then with the same $T_{nw}$ and higher $\lambda_{nw}$, $T_w$ has to decrease to maintain the same heat flux (it is similar in all the numerical setup for a given case). This explains the high sensitivity of wall temperature maximum to the smallest roughness presence.

Although this is not sufficient to explain the shift in the onset location of HTD, the cause for the delay is the direct consequence of lower wall temperature. The primary reason for the downstream shift of the HTD onset location is the weakening of dominant forces (buoyancy for case-2 & acceleration effects for case-3) responsible for HTD. In other words, apart from increasing the turbulence level, roughness also mitigates the forces accountable for a decline in TKE and eventually HTD. The same can be explained by the fact that all these forces emanate from density variation, which results from temperature variation as the former vary monotonically with the latter. Moreover, as concluded above, $T_w$ will decrease with roughness presence, and the same will reduce the temperature disparity (as well as density variation) in the flow. It will ultimately lead to the weakening of dominant forces (buoyancy or acceleration), and the same can be demonstrated by enumerating the magnitude of these forces based on equations (14 &15) as proposed by the author[24]. Figure 11(b) displays the difference of both the forces for case-2, and as mentioned above, the buoyancy forces are decreasing with increases in roughness. As a result, the difference between the two forces is less for a given axial location for higher $K^*_s$. Consequently, the balance between both forces exists for a longer axial distance. It eventually restricts the decline of TKE that occurs in usual HTD. So, delay in the HTD onset is a cascading effect of the increased turbulence level followed by weakening of dominant forces where the former is stimulated from the latter.

$$\Delta p_b \sim (\rho_{co} - \rho_{an}) g \Delta x \left(\frac{CA_{\Delta r}}{CA}\right) \tag{14}$$

$$\Delta p_a \sim G^2 \left(\frac{\overline{\rho}_{x_1} - \overline{\rho}_{x_2}}{\overline{\rho}_{x_1}\overline{\rho}_{x_2}}\right) \tag{15}$$

It is also worth pointing that for $K^*_{st} \geq 37.5 \times 10^{-4}$, the balance between two forces is throughout the flow domain as shown in figure 11(b) for case-2; thus, HTD does not occur, as highlighted in the author's previous work[24]. The same can be referred for the nomenclature used in the equation (14 &15), as the above equations are provided here for completeness.

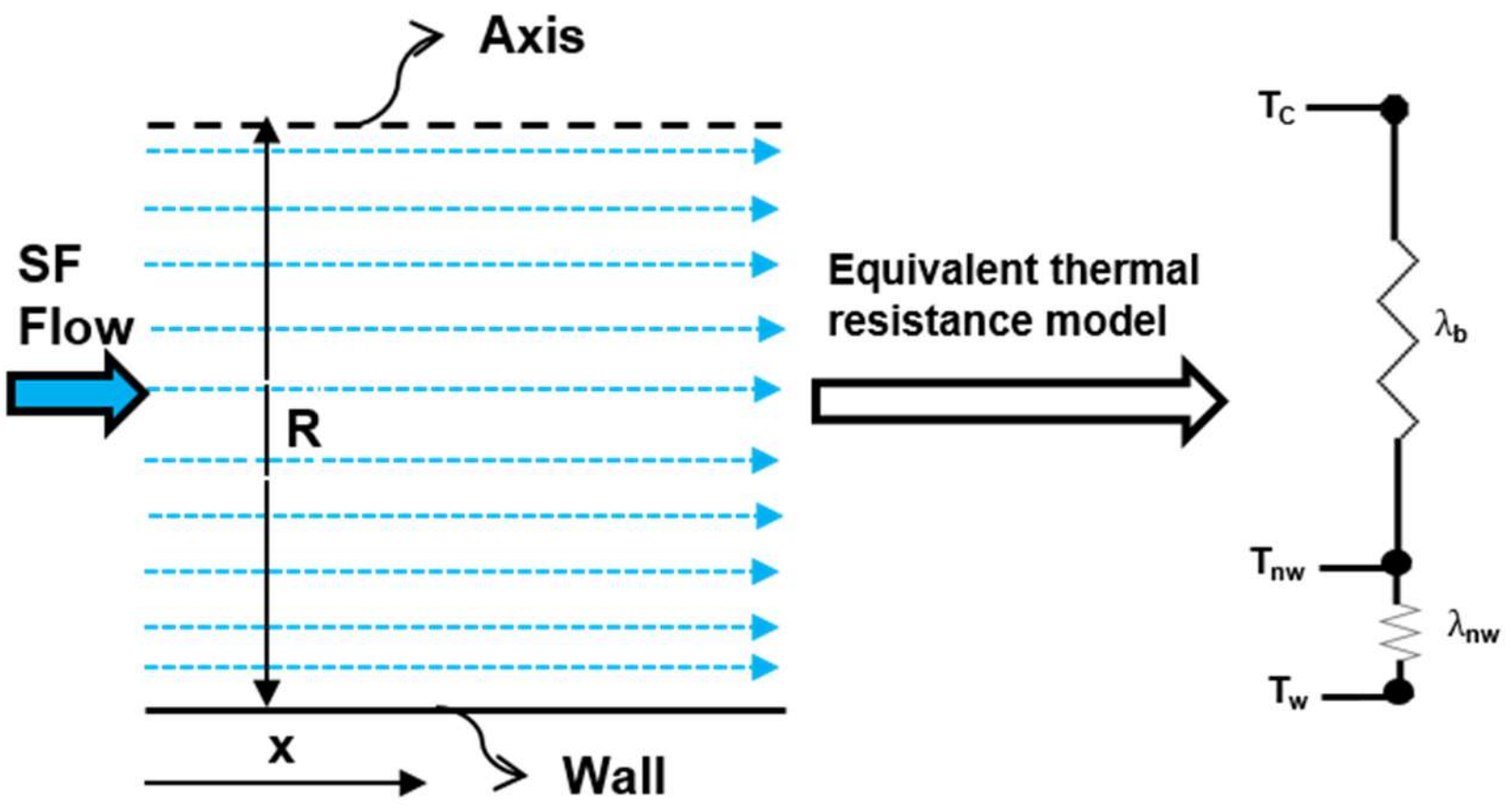

**(a)**

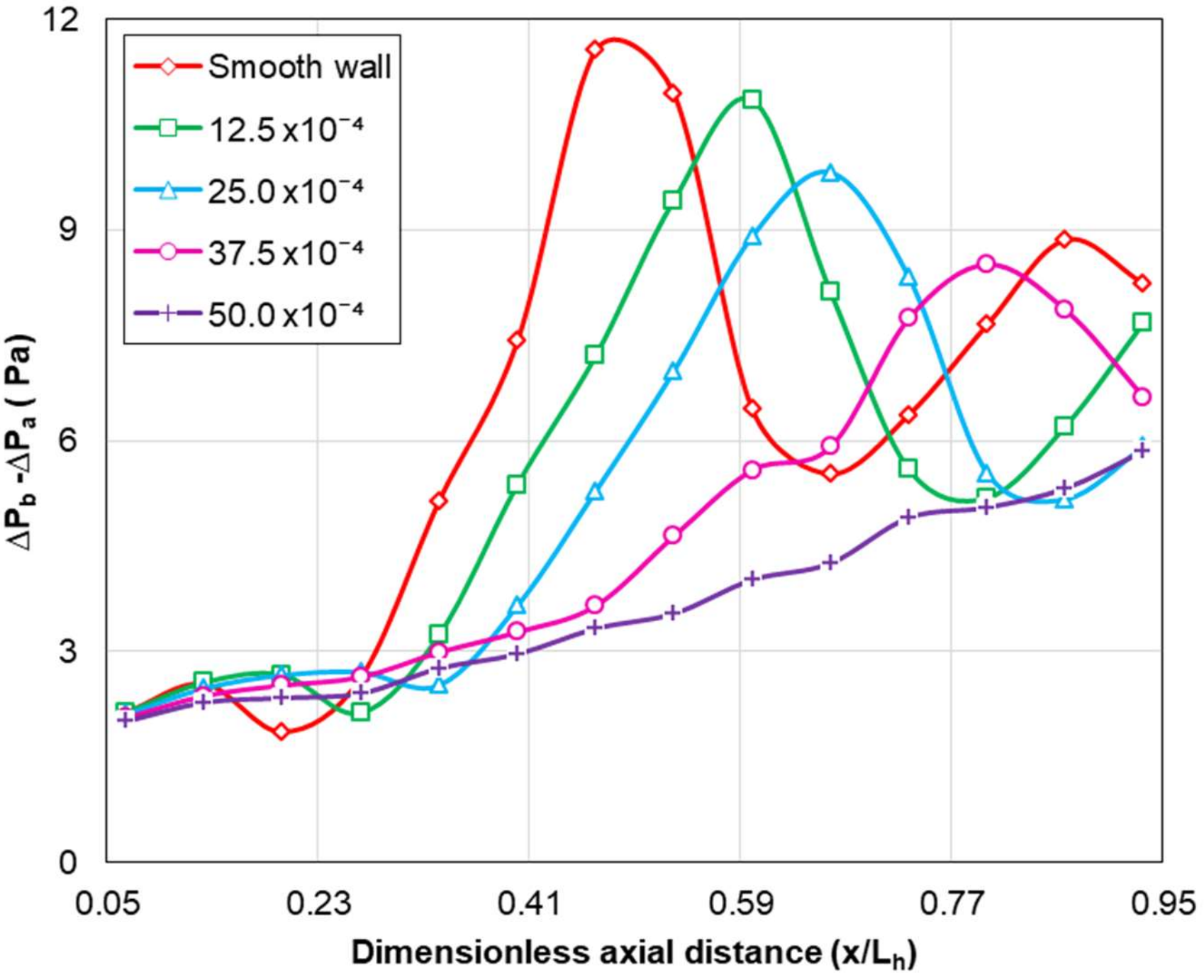

**(b)**

**Figure 11. (a) Schematic showing the equivalent thermal resistance model for SF flow (b) Plot showing the difference of buoyancy and acceleration forces ($\Delta p_b - \Delta p_a$) based on equation (14&15) for case-2. Numbers in the legend show the $K_s^*$ magnitude.**

### 3.3.2 A new form of heat transfer impairment ($HTD_n$) near the outlet

Figure 9(b) in section (3.2.2) has unveiled a new form of heat transfer impairment ($HTD_n$), where unlike usual, HTD loss of turbulence is not the cause. In other words, $HTD_n$ is not the output of flow field manipulation that results in the decline of TKE, whereas the same leads to typical HTD occurrence reported in the SF flow. It is bolstered by the fact that TKE (as shown in figure 10(b)) is augmented as $K_s^*$ value is increased, and surprisingly $HTD_n$ has been evident for the roughness magnitudes ($K_s^* \geq 40$ x$10^{-4}$) in which the usual HTD has completely recovered.

The most probable reason for $HTD_n$ seems to be associated with the higher specific heat value at the pseudocritical temperature ($T_{pc}$). Authors [24] have shown pseudo-boiling presence in SF flow heat transfer because a higher specific heat region may act as a saturation (two-phase region) zone in subcritical phase change. It absorbs energy with no change in temperature and eventually results in increased thermal resistance in the flow; for example, it will decrease $\lambda_b$ & $\lambda_{nw}$ shown in figure 11(a).

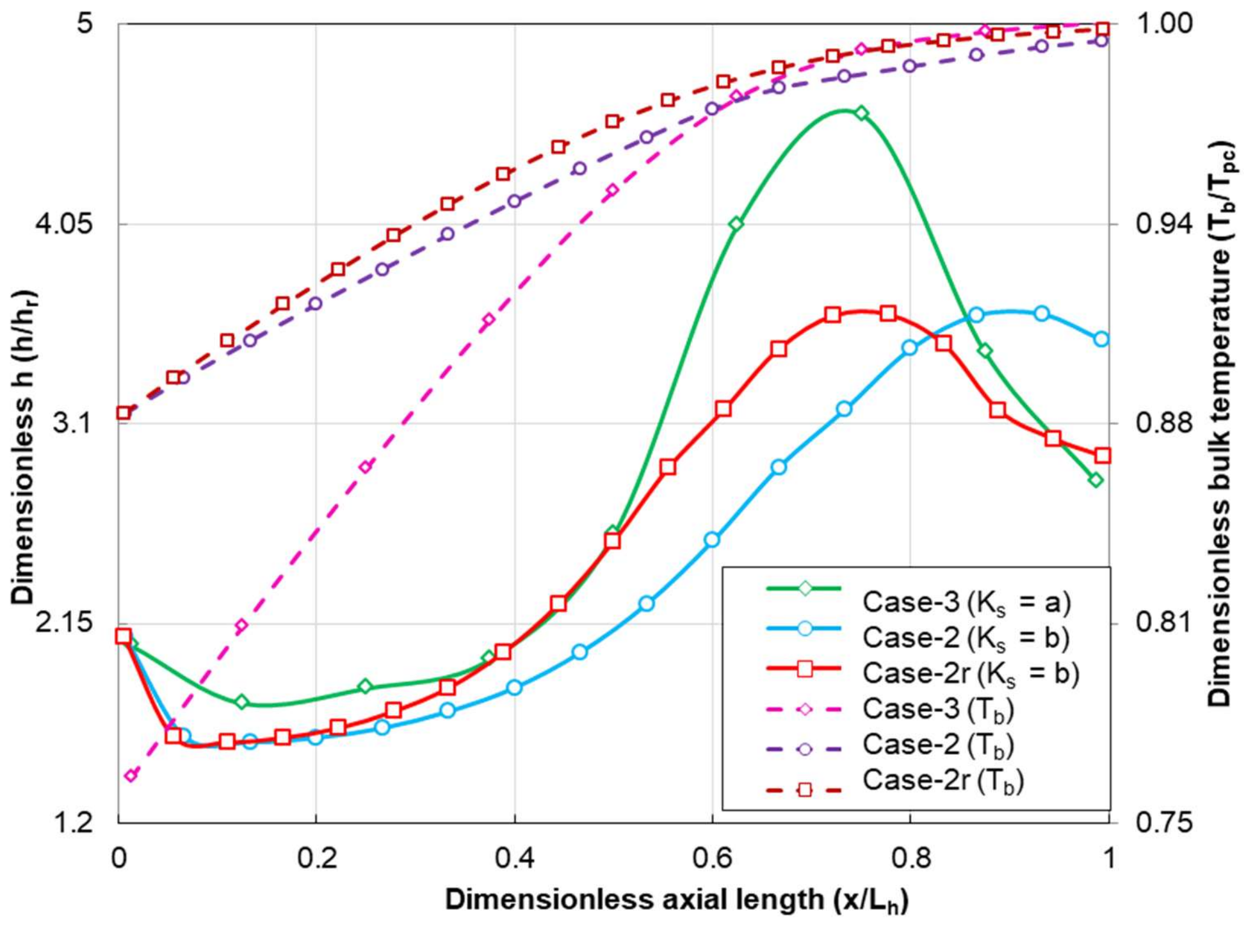

**(a) Value of variables in the legend is: a = 8µm and b= 20µm**

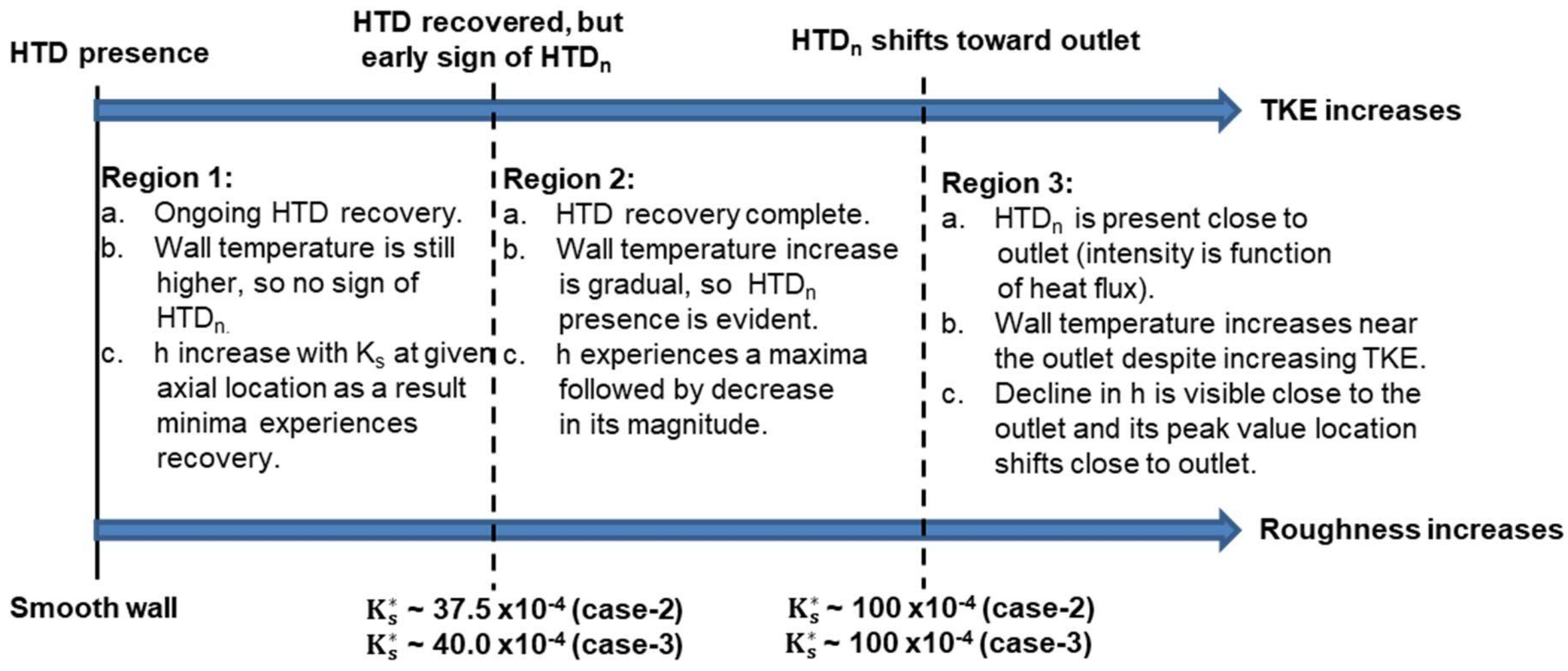

**(b)**

**Figure 12. (a) Plot showing the axial variation of dimensionless heat transfer coefficient and bulk fluid temperature for setup mentioned in the legend. Besides, $h_r$ used to normalize the h is the average value of h for smooth pipe for a given case. (b) Schematic summarising the flow regime of case-2 &3 in three regions with respect to HTD and $HTD_n$ observation.**

Furthermore, it was observed that the bulk temperature ($T_b$) in case-3 approaches the $T_{pc,}$ opposed to case-2, as shown in figure 12(a). It essentially means that the significant chunk of fluid near the outlet is closer to $T_{pc,}$ thus having a very high specific heat value. So, to maintain the same heat flux, wall temperature will rise to compensate for dwindling values of $K_{nw}$ and $K_b$ in the flow domain. It is worth pointing that the $HTD_n$ location shifts closer to the outlet with increasing $K_s$ value. For example, for $K_s^* \geq 40$ x$10^{-4}$, h has a peak at x=0.5m, whereas it is close to x=0.6m and x=0.7m for $K_s^*$ value of 53.3 x$10^{-4}$ and 100 x$10^{-4}$, respectively. It can be attributed to the fact that the TKE value increases with $K_s$ and, consequently, thermal conductance ($\lambda_{nw}$ & $\lambda_b$). Hence for the same heat flux, the increase in wall temperature will be early for lower $K_s$ value, and the same will govern the location of h maximum.

Although case-3 has one order higher magnitude of TKE than case-2, $HTD_n$ occurs in former instead of later. So, in order to bolster the explanation discussed above, a modified flow domain (case-2r) was created by increasing the length of $L_h$ in case-2 by 0.3m so that $T_b$ at the outlet moves closer to $T_{pc}$. Surprisingly, It was observed that $HTD_n$ does occur in case-2r (Case-2 with revised $L_h$ value of 1.8m) and is more evident than case-2 for the same flow condition ($K_s$=20μm) as displayed in figure 12(a). All the axis of the plot is in the dimensionless form to accommodate all the results with better clarity and perspective. Interestingly, the trend observed for each case is similar in dimensionless form and the same points toward the universal nature of SF flow with HTDn. Besides, $HTD_n$ severity is a direct consequence of the heat flux applied at the wall. For example, case-3 experiences a drastic decrease in heat transfer coefficient compared to case-2 as the former has a higher heat flux value than the latter. The same has been highlighted in figure 12(b), which represents a schematic that tries to characterize the SF flow with roughness presence into three broad regions based on the HTD and $HTD_n$ occurrence. This demarcation is based on the $K_s^*$ for which similar observations were made in both the cases (2 &3), although the $HTD_n$ magnitude was different.

### 3.4 Efficacy analysis of roughness presence

#### 3.4.1 Second law Analysis

Average entropy generation ($\overline{S}$) in the flow domain has been used as a quantitative parameter for second law analysis, and it has been calculated by performing the volume integral of the equation (8-11) over the flow domain as shown in equation (16&17). Besides, to make this analysis more inclusive, case-1 (as given in Table 1) has been investigated along with case-2 & 3. All these entropy generation terms have been plotted as a function of dimensionless roughness height ($K_s^*$) as displayed in Figure 13 (a) and (b).

$$\overline{S}_{U1} = \iiint \dot{S}_{U1} \;\; \& \, \overline{S}_{U2} = \iiint \dot{S}_{U2} \tag{16}$$

Similarly,

$$\overline{S}_{T1} = \iiint \dot{S}_{T1} \;\& \, \overline{S}_{T2} = \iiint \dot{S}_{T2} \tag{17}$$

Moreover, figure 13(a) displays the viscous entropy generation term ($\overline{S}_U$), and interestingly all the cases follow a similar trend, although the magnitude is one order higher for case-3 compared to case-1&2. It can be explained based on the higher mass flux in case-3. Initially, $\overline{S}_{U1}$ term grows slightly as roughness increases, but the rise is not rapid as in the case of $\overline{S}_{U2}$ generation term. This trend is on the expected lines as turbulence levels go up with higher roughness. However, $\overline{S}_{U1}$ experiences a dip after being more or less constant, and then it gradually recovers. This decline may be attributed to the fact that for rough pipe velocity profile near the wall becomes less steep compared to a smooth pipe. On the contrary, $\overline{S}_{U2}$ grows with increasing $K_s^*$ as it depends on turbulence level.

Further, $\overline{S}_{T1}$ and $\overline{S}_{T2}$ representing entropy generation due to thermal gradients has a maximum for some intermediate value of roughness ($K_{sc}$ or $K^*_{sc}$) and as roughness increases, its value decreases and follows smooth variation. This peculiar behavior has been observed for all the cases: $K^*_{sc}$ is 25 $\times10^{-4}$ for both case-1&2 whereas it is 26.6 $\times10^{-4}$ for case-3. However, $\overline{S}_{T2}$ value is higher than that of $\overline{S}_{T1}$ for all the cases, which is expected because the former one is due to turbulent variation. Besides, in all the cases $\overline{S}_{U}$ ($\overline{S}_{U1}$ or $\overline{S}_{U2}$) is much less than the $\overline{S}_{T}$ ($\overline{S}_{T1}$ or $\overline{S}_{T2}$): there is a difference of order $10^{-5}$ for case-1&2, whereas it is $10^{-4}$ for case-3. As a result, the total entropy generation ($\overline{S}_{Total}$) defined in equation (18) is approximately equal to the sum of $\overline{S}_{T1}$ and $\overline{S}_{T2}$ and it follows the same trend displayed individually by $\overline{S}_{T}$ rather than $\overline{S}_{U}$ term.

$$\overline{S}_{Total} = \overline{S}_{U1} + \overline{S}_{U2} + \overline{S}_{T1} + \overline{S}_{T2} \tag{18}$$

The peak in entropy generation can be interpreted by tracing the variables on which $\overline{S}_{T}$ depends. It is a function of: (I) Temperature gradient $\left(\frac{\partial T}{\partial x} \& \frac{\partial T}{\partial r}\right)$: The temperature derivative will be higher for the SF flow with HTD occurrence as shown in figure 6(a). The same will hold true for all the $K^*_{s}$ values where HTD have not recuperated completely. Interestingly, HTD still exist for $K^*_{sc}$ in all the cases as displayed in figure 7(a) & 9(a). However, once HTD has recovered completely, the thermal gradient value will reduce drastically. (II) Temperature (T): Similar to the temperature gradient, temperature will be higher at a given location for the flow condition encountering HTD, and its value will reduce after HTD has vanished. (III) Thermal conductivity (λ): Its value decreases with an increase in temperature or vice-versa, as shown in figure 2. It will increase continuously in the flow domain with $K^*_{s}$. So $\overline{S}_{T}$ will have maximum value when (I) and (III) are on the higher side, whereas (II) is on the lower side.

There is a significantly less difference in the $\overline{S}_{Total}$ magnitude for roughness less than $K^*_{sc}$, as no significant impact of roughness is observed on the bulk of the flow. The most probable reason for the peak in $\overline{S}_{Total}$ is the increased value of λ due to lower temperature as the severity of HTD attenuates with $K^*_{s}$. Since the change in other variables (I &II) governing the $\overline{S}_{T}$ will compensate each other for $K^*_{s} \leq K^*_{sc}$. In other words, a lower temperature will be negated by decreased thermal gradient value as they form a ratio term in $\overline{S}_{T}$. Besides, after HTD has been mitigated completely for roughness beyond the $K^*_{sc}$, thermal gradient decreases substantially. Consequently $\overline{S}_{Total}$ suffers a steep decline as shown in figure 13(b) despite having a high value of λ. It is worth highlighting that the maxima value is not that apparent for case-1 compared to other cases because of no HTD occurrence and the $\overline{S}_{Total}$ is highest for case-3 for all $K^*_{sc}$ owing to its higher heat flux.

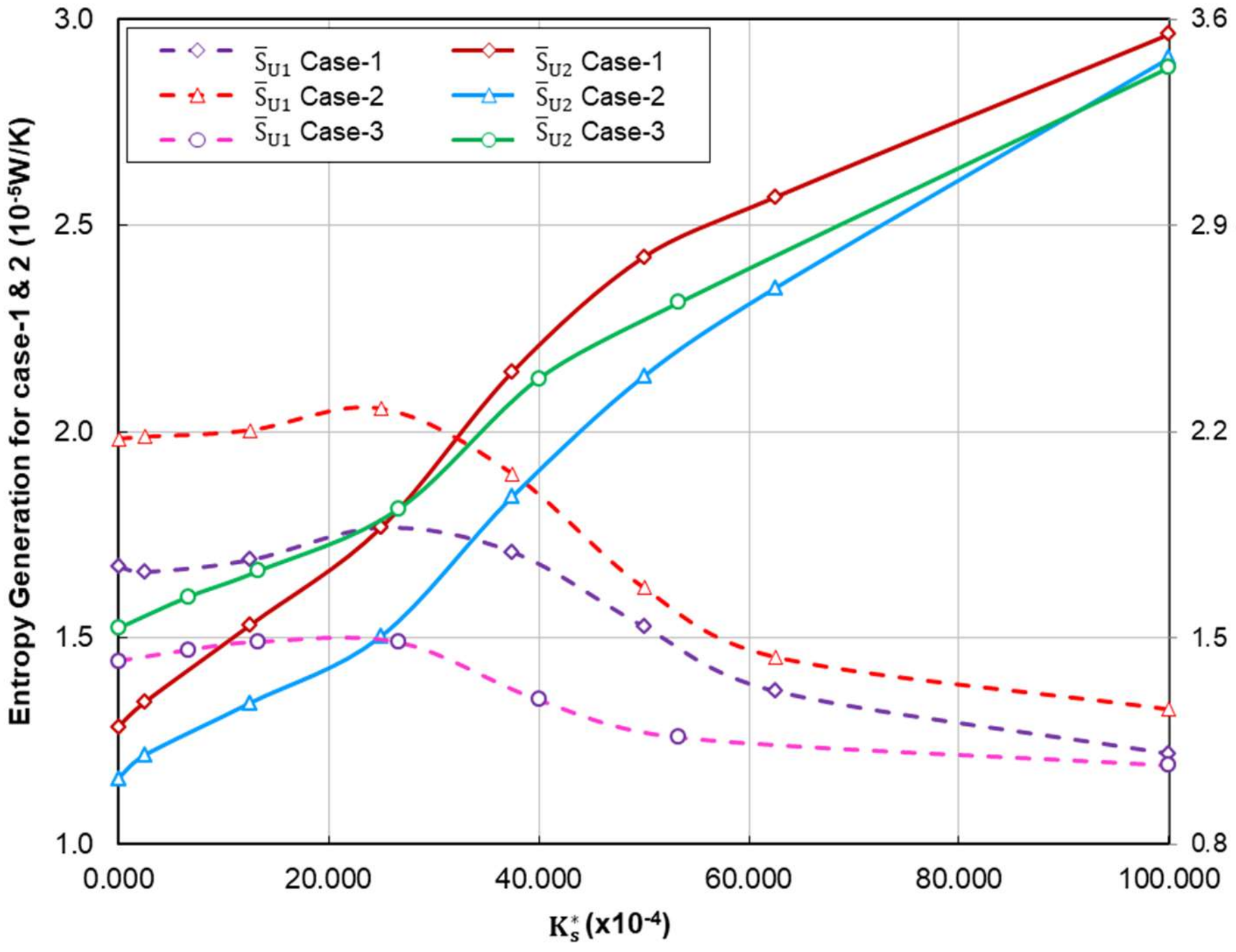

**(a)**

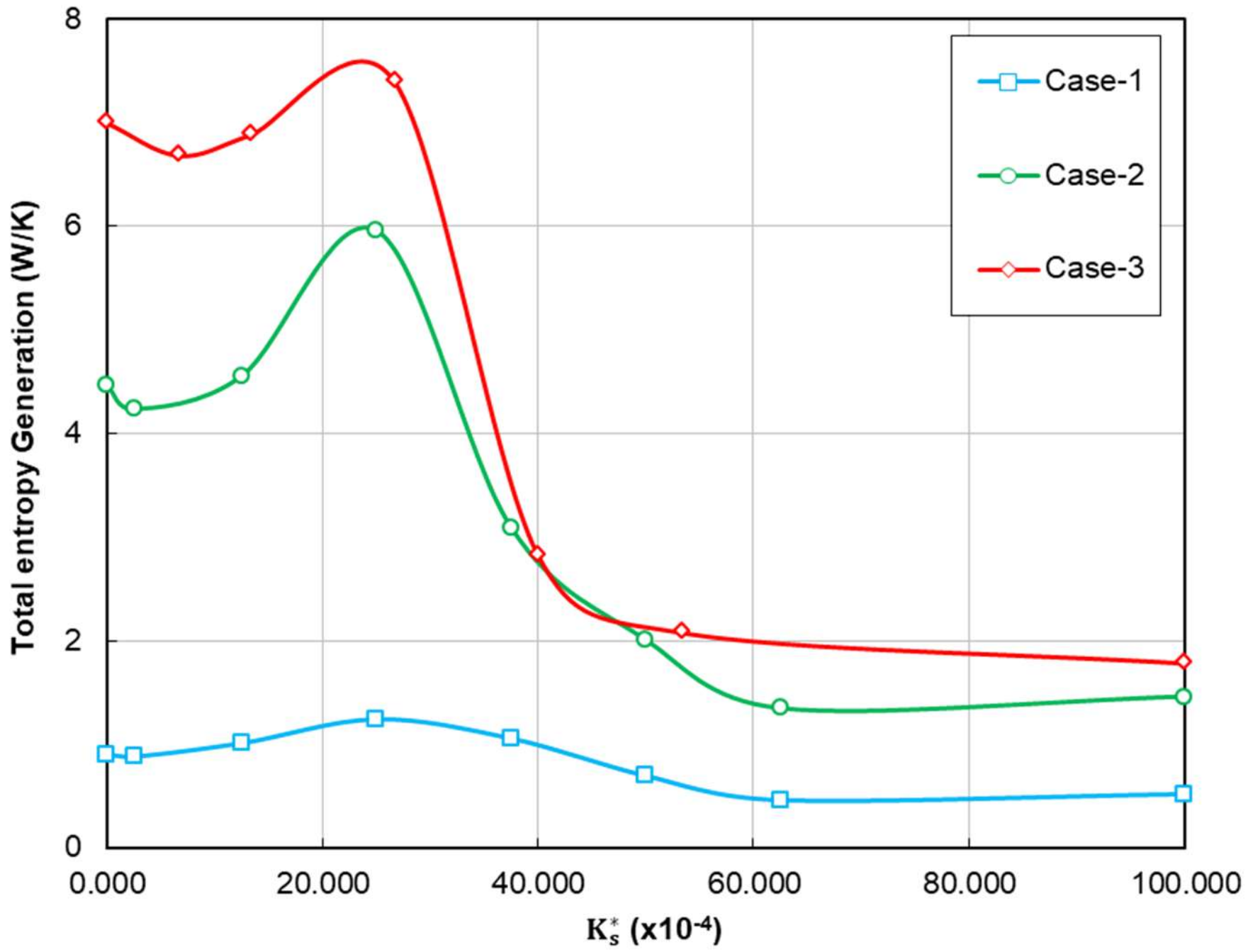

**(b)**

**Figure 13. (a) Entropy generation due to dissipation ($\overline{S}_U$) variation with dimensionless roughness height ($K_s^*$) for all the cases. (b) Entropy generation due to thermal gradient ($\overline{S}_T$) variation with dimensionless roughness height ($K_s^*$) for all the cases.**

### 3.4.2 Thermal factor analysis

Roughness presence is not advantageous in all aspects of convective heat transfer. For example, although it augments the heat transfer by adding turbulence to the flow, the pressure loss incurred during the flow also increases drastically, which directly means more pumping power. So, thermal factor analysis (η) described in equation (6) judges the efficacy of roughness by considering these contrasting aspects of roughness and present it in the form of ration. The equation (6) can be further manipulated based on the following definition based on average quantities:

$$\mathrm{Nu} = \frac{\overline{h}D}{\lambda} \;\&\; f = \frac{\tau}{\rho U^2/2} \sim \frac{(\Delta P/L)D}{\rho U^2/2} \tag{19}$$

As for a given case (1,2 &3), flow variables and properties can be approximated to be of similar magnitude because of the same flow conditions. So, using equation (19), equation (6) can be redefined for a given case as following:

$$\eta = \frac{\overline{h}/\overline{h}_r}{(\Delta p/\Delta p_r)^{1/3}} \tag{20}$$

Here 'r' subscript denotes the reference values that are taken equal to the smooth pipe results for a given case. Also, $\overline{h}$ represents average heat transfer coefficient and $\Delta p$ stands for pressure loss incurred during the flow. Figure (14) shows the η variation with Ks/R for all the cases. Contrary to second law analysis, this efficacy study suggests that increasing roughness is always beneficial. There is no threshold value ($K_{sc}$) similar to section (3.4.1) beyond which one should select the $K_s$ value. For all the cases, a subsequent increase of roughness results in higher heat transfer gain than the rise in pumping power employed to counter the shear stress presented at the wall. However, case-1 achieves a break-even kind of situation at $K_{sc}^* = 40 \times 10^{-4}$, where the augmentation in heat transfer is the same as increased pressure loss. It can be anticipated because, unlike case-2 &3, it does not experience any HTD, so adding roughness will not improve the heat transfer significantly, although it will increase the pumping power. As a result, η will not experience any considerable growth and will maintain approximately at a constant value. Also, it might be possible that similar behavior can be displayed in case-2 &3 at a higher roughness value.

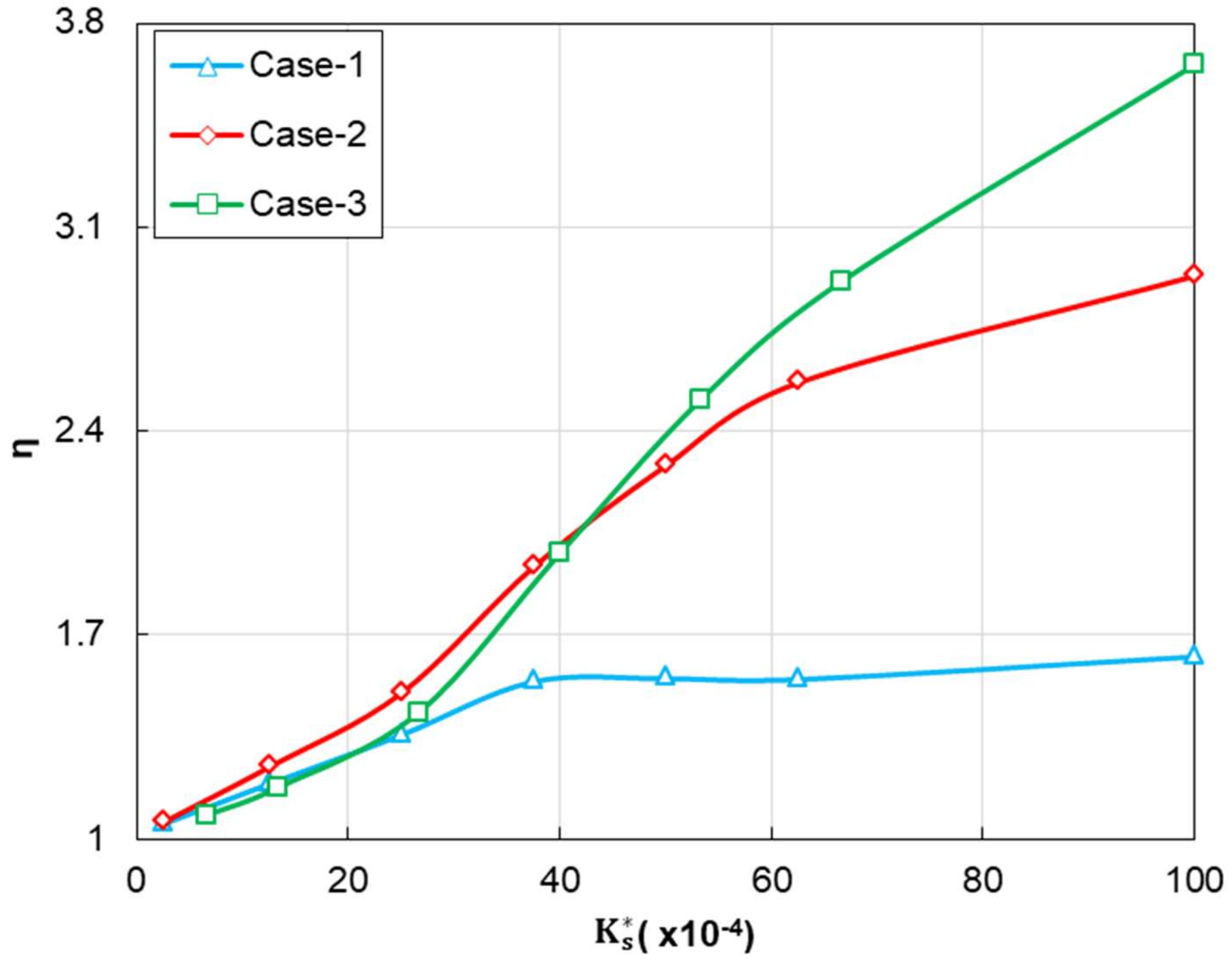

**Figure 14. Thermal performance factor for different cases as a function of $K_s^*$.**

## 4. CONCLUSION

The present study investigates the supercritical water flow in a rough circular pipe. The numerical setup uses a roughness model that is based on the Nikuradse experiment. As expected, HTD recovers with the increase in roughness for all the cases. Although the bulk of fluid is influenced only after $K_{st}$, the wall temperature peaks are very sensitive to the presence of even the smallest roughness height. Besides, the onset of HTD is delayed and shifted further downstream with increasing $K_s$. It is a combined effect of increased turbulence and decline of dominant force term governing the flow.

Furthermore, a new type of heat transfer impairment (HTDn) different from the usual HTD is observed, and the reasoning lies in the higher specific heat value at $T_{pc.}$ Finally, roughness effectiveness is judged using the second law analysis and thermal performance factor (η). The former reveals a $K_{sc}$ value that should be avoided as maximum entropy generation occurs for the same. In contrast, no such conclusion can be drawn from the later analysis as η increases with $K_s$. It is found that the variation in thermal conductivity with temperature is responsible for the maximum entropy generation. This study will encourage experimentalists to report the surface texture of the test section apart from the usual flow condition, especially for the SF flows encountering HTD.

Nevertheless, this paper explores the roughness response on the supercritical flow; there is a need for experiments to study the same as it will provide greater insight into flow behavior and help us generate the roughness model for supercritical flow. On the numerical front, the inclusion of roughness is immensely challenging; for example, while the roughness model (based on the Nikuradse experiment) for RANS simulation, the value of the $C_s$ is not clearly defined. This model is suitable for particular roughness characteristics such as uniform and tightly packed protrusion. There is a need for detailed and extensive numerical methods such as direct numerical simulations (DNS) for treating the surface roughness in supercritical flow. Together with experiments, DNS can contribute to a robust and generic roughness model that can incorporate all kinds of roughness structure and distribution. In the end, $HTD_n$ can be investigated further with an operating condition such that bulk temperature approaches pseudocritical point.

**Acknowledgment**

The authors will like to acknowledge the financial support from the Department of science and technology (DST), Government of India under the National Center for Clean Coal Research and development (NCCCR&D) scheme.